\documentclass[twocolumn, twocolappendix]{aastex631}

\definecolor{Red}{rgb}{0.9,0,0}
\definecolor{Blue}{rgb}{0,0,0.9}
\definecolor{Green}{rgb}{0,0.5,0}
\definecolor{Black}{rgb}{0,0,0}

\usepackage{color, colortbl, soul}
\usepackage{amsmath}
\usepackage{subfigure}

\newcommand{\Autoref}[1]{%
  \begingroup%
  \def\chapterautorefname{Chapter}%
  \def\sectionautorefname{Section}%
  \def\subsectionautorefname{Subsection}%
  \def\subsubsectionautorefname{Subsubsection}%
  \def\paragraphautorefname{Paragraph}%
  \def\tableautorefname{Table}%
  \def\equationautorefname{Equation}%
  \autoref{#1}%
  \endgroup%
}

\received{January 20, 2022}
\revised{March 8, 2022}
\accepted{March 14, 2022}
\submitjournal{AJ}

\shorttitle{Stellar occultation by the Galilean moons}
\shortauthors{Morgado et al.}

\begin{document}

\title{Milliarcsecond astrometry for the Galilean moons using stellar occultations}

\author{B. E. Morgado}
\affiliation{Observatório Nacional/MCTI, R. General José Cristino 77, CEP 20921-400 Rio de Janeiro - RJ, Brazil}
\affiliation{Laboratório Interinstitucional de e-Astronomia - LIneA, Rua Gal. José Cristino 77, Rio de Janeiro, RJ 20921-400, Brazil}
\affiliation{LESIA, Observatoire de Paris, Université PSL, CNRS, Sorbonne Université, Univ. Paris Diderot, Sorbonne Paris Cité, 5 place Jules Janssen, 92195 Meudon, France}
\affiliation{Universidade Federal do Rio de Janeiro - Observatório do Valongo, Ladeira Pedro Antônio 43, CEP 20.080-090 Rio de Janeiro - RJ, Brazil}
\correspondingauthor{B. E. Morgado}
\email{morgado.fis@gmail.com}

\author{A. R. Gomes-Júnior}
\affiliation{UNESP - São Paulo State University, Grupo de Dinâmica Orbital e Planetologia, CEP 12516-410, Guaratinguetá, SP, Brazil}
\affiliation{Laboratório Interinstitucional de e-Astronomia - LIneA, Rua Gal. José Cristino 77, Rio de Janeiro, RJ 20921-400, Brazil}

\author{F. Braga-Ribas}
\affiliation{Federal University of Technology - Paraná (UTFPR / DAFIS), Rua Sete de Setembro, 3165, CEP 80230-901, Curitiba, PR, Brazil}
\affiliation{Observatório Nacional/MCTI, R. General José Cristino 77, CEP 20921-400 Rio de Janeiro - RJ, Brazil}
\affiliation{Laboratório Interinstitucional de e-Astronomia - LIneA, Rua Gal. José Cristino 77, Rio de Janeiro, RJ 20921-400, Brazil}
\affiliation{LESIA, Observatoire de Paris, Université PSL, CNRS, Sorbonne Université, Univ. Paris Diderot, Sorbonne Paris Cité, 5 place Jules Janssen, 92195 Meudon, France}

\author{R. Vieira-Martins}
\affiliation{Observatório Nacional/MCTI, R. General José Cristino 77, CEP 20921-400 Rio de Janeiro - RJ, Brazil}
\affiliation{Laboratório Interinstitucional de e-Astronomia - LIneA, Rua Gal. José Cristino 77, Rio de Janeiro, RJ 20921-400, Brazil}
\affiliation{Universidade Federal do Rio de Janeiro - Observatório do Valongo, Ladeira Pedro Antônio 43, CEP 20.080-090 Rio de Janeiro - RJ, Brazil}

\author{J. Desmars}
\affiliation{Institut Polytechnique des Sciences Avancées IPSA, 63 boulevard de Brandebourg, 94200 Ivry-sur-Seine, France}
\affiliation{IMCCE, Observatoire de Paris, PSL University, CNRS, Sorbonne Université, Université Lille, 77 Av. Denfert-Rochereau, 75014 Paris, France}

\author{V. Lainey}
\affiliation{IMCCE, Observatoire de Paris, PSL University, CNRS, Sorbonne Université, Université Lille, 77 Av. Denfert-Rochereau, 75014 Paris, France}

\author{E. D'aversa}
\affiliation{Istituto Nazionale di Astrofisica-Istituto di Astrofisica e Planetologia Spaziali, INAF-IAPS, Rome, Italy}

\author{D. Dunham}
\affiliation{International Occultation Timing Association (IOTA), PO Box 20313, Fountain Hills, AZ 85269, USA}

\author{J. Moore}
\affiliation{International Occultation Timing Association (IOTA), PO Box 20313, Fountain Hills, AZ 85269, USA}

\author{K. Baillié}
\affiliation{IMCCE, Observatoire de Paris, PSL University, CNRS, Sorbonne Université, Université Lille, 77 Av. Denfert-Rochereau, 75014 Paris, France}

\author{D. Herald}
\affiliation{International Occultation Timing Association / European Section, Am Brombeerhag 13, D-30459 Hannover, Germany}
\affiliation{Trans-Tasman Occultation Alliance (TTOA), Wellington PO Box 3181, New Zealand}

\author{M. Assafin}
\affiliation{Universidade Federal do Rio de Janeiro - Observatório do Valongo, Ladeira Pedro Antônio 43, CEP 20.080-090 Rio de Janeiro - RJ, Brazil}
\affiliation{Laboratório Interinstitucional de e-Astronomia - LIneA, Rua Gal. José Cristino 77, Rio de Janeiro, RJ 20921-400, Brazil}

\author{B. Sicardy}
\affiliation{LESIA, Observatoire de Paris, Université PSL, CNRS, Sorbonne Université, Univ. Paris Diderot, Sorbonne Paris Cité, 5 place Jules Janssen, 92195 Meudon, France}

\author{S. Aoki}
\affiliation{Japan Aerospace Exploration Agency, JAXA}

\author{J. Bardecker}
\affiliation{International Occultation Timing Association (IOTA), PO Box 20313, Fountain Hills, AZ 85269, USA}

\author{J. Barton}
\affiliation{International Occultation Timing Association (IOTA), PO Box 20313, Fountain Hills, AZ 85269, USA}

\author{T. Blank}
\affiliation{International Occultation Timing Association (IOTA), PO Box 20313, Fountain Hills, AZ 85269, USA}

\author{D. Bruns}
\affiliation{International Occultation Timing Association (IOTA), PO Box 20313, Fountain Hills, AZ 85269, USA}

\author{N. Carlson}
\affiliation{International Occultation Timing Association (IOTA), PO Box 20313, Fountain Hills, AZ 85269, USA}

\author{R. W. Carlson}
\affiliation{Jet Propulsion Laboratory, California Institute of Technology, Pasadena, CA, USA}

\author{K. Cobble}
\affiliation{International Occultation Timing Association (IOTA), PO Box 20313, Fountain Hills, AZ 85269, USA}

\author{J. Dunham}
\affiliation{International Occultation Timing Association (IOTA), PO Box 20313, Fountain Hills, AZ 85269, USA}

\author{D. Eisfeldt}
\affiliation{International Occultation Timing Association (IOTA), PO Box 20313, Fountain Hills, AZ 85269, USA}

\author{M. Emilio}
\affiliation{Universidade Estadual de Ponta Grossa (UEPG), Ponta Grossa, Brazil}

\author{C. Jacques}
\affiliation{Observatório SONEAR, Brazil}

\author{T. C. Hinse}
\affiliation{Institute of Astronomy, Faculty of Physics, Astronomy and Informatics, Nicolaus Copernicus University, Grudziadzka 5, 87-100, Torun, Poland}
\affiliation{Chungnam National University, Department of Astronomy and Space Science, 34134, Daejeon, Republic of Korea}

\author{Y. Kim}
\affiliation{Department of Astronomy and Space Science, Chungbuk National University, 28644 Cheongju, South Korea}
\affiliation{Chungbuk National University Observatory, Chungbuk National University, 28644 Cheongju, South Korea}

\author{M. Malacarne}
\affiliation{Universidade Federal do Espírito Santo, Av. Fernando Ferrari 514, Vitória, ES, 29075-910, Brazil}

\author{P. D. Maley}
\affiliation{International Occultation Timing Association (IOTA), PO Box 20313, Fountain Hills, AZ 85269, USA}
\affiliation{NASA Johnson Space Center Astronomical Society, Houston, TX, USA}

\author{A. Maury}
\affiliation{San Pedro de Atacama Celestial Explorations - SPACE, Chile}

\author{E. Meza}
\affiliation{Comisión Nacional de Investigación y Desarrollo Aeroespacial del Perú - CONIDA}
\affiliation{Observatorio Astronómico de Moquegua - MPC Code W73}

\author{F. Oliva}
\affiliation{Istituto Nazionale di Astrofisica-Istituto di Astrofisica e Planetologia Spaziali, INAF-IAPS, Rome, Italy}

\author{G. S. Orton}
\affiliation{Jet Propulsion Laboratory, California Institute of Technology, Pasadena, CA, USA}

\author{C. L. Pereira}
\affiliation{Federal University of Technology - Paraná (UTFPR / DAFIS), Rua Sete de Setembro, 3165, CEP 80230-901, Curitiba, PR, Brazil}
\affiliation{Observatório Nacional/MCTI, R. General José Cristino 77, CEP 20921-400 Rio de Janeiro - RJ, Brazil}
\affiliation{Laboratório Interinstitucional de e-Astronomia - LIneA, Rua Gal. José Cristino 77, Rio de Janeiro, RJ 20921-400, Brazil}

\author{M. Person}
\affiliation{MIT, Cambridge, MA, USA}

\author{C. Plainaki}
\affiliation{Agenzia Spaziale Italiana, ASI, Rome, Italy}

\author{R. Sfair}
\affiliation{UNESP - São Paulo State University, Grupo de Dinâmica Orbital e Planetologia, CEP 12516-410, Guaratinguetá, SP, Brazil}
\affiliation{Institut für Astronomie und Astrophysik, Eberhard Karls Universität Tübingen, Auf der Morgenstelle 10, 72076 Tübingen, Germany}

\author{G. Sindoni}
\affiliation{Agenzia Spaziale Italiana, ASI, Rome, Italy}

\author{M. Smith}
\affiliation{International Occultation Timing Association (IOTA), PO Box 20313, Fountain Hills, AZ 85269, USA}

\author{E. Sussenbach}
\affiliation{Pletterijweg Oost, Willemstad, Curaçao}

\author{P. Stuart}
\affiliation{International Occultation Timing Association (IOTA), PO Box 20313, Fountain Hills, AZ 85269, USA}

\author{J. Vrolijk}
\affiliation{Space and Nature Aruba Foundation, Aruba}

\author{O. C. Winter}
\affiliation{UNESP - São Paulo State University, Grupo de Dinâmica Orbital e Planetologia, CEP 12516-410, Guaratinguetá, SP, Brazil}



\begin{abstract}

A stellar occultation occurs when a Solar System object passes in front of a star for an observer. This technique allows the determination of sizes and shapes of the occulting body with kilometer precision. Also, this technique constrains the occulting body's positions, albedos, densities, etc. In the context of the Galilean moons, these events can provide their best ground-based astrometry, with uncertainties in the order of 1 mas ($\sim$ 3 km at Jupiter's distance during opposition). We organized campaigns and successfully observed a stellar occultation by Io (JI) in 2021, one by Ganymede (JIII) in 2020, and one by Europa (JII) in 2019, with stations in North and South America. Also, we re-analyzed two previously published events, one by Europa in 2016 and another by Ganymede in 2017. Then, we fit the known 3D shape of the occulting satellite and determine its center of figure. That resulted in astrometric positions with uncertainties in the milliarcsecond level. The positions obtained from these stellar occultations can be used together with dynamical models to ensure highly accurate orbits of the Galilean moons. These orbits can help plan future space probes aiming at the Jovian system, such as JUICE by ESA and Europa Clipper by NASA, and allow more efficient planning of flyby maneuvers.

\end{abstract}



\section{Introduction} \label{sec:intro}

The progress in astrometry and modeling the orbits of planetary satellites in the last decade made possible the accurate estimation of tidal effects in natural satellites and their primaries \citep{Lainey_2009, Lainey_2012, Lainey_2017}. These studies need accurate observations spread over a significant period of time to provide essential constraints on short and long-term dynamics, up to formation processes \citep{Charnoz_2011, Crida_2012}.

Accurate orbits also help prepare space missions targeting these systems \citep{Dirkx_2016, Dirkx_2017}. The ESA -- JUICE and the NASA -- Europa Clipper missions are scheduled to be launched this decade (the 2020s) for the Jovian system. The accurate ground-based astrometry of natural satellites is important for these studies.

In the CCD era, straightforward classical astrometry can not be used for all the natural satellites. The case of Galilean moons is a critical example since Jupiter's brightness in the Field of View (FoV) would quickly saturate the CCD, thus providing positions with uncertainties that range between 100 and 150 milliarcseconds (mas) \citep{Kiseleva_2008}. This scenario motivates the search for alternative methods for the astrometry of these satellites, for example, the mutual phenomena events \citep[and references therein]{Aksnes_1976, Aksnes_1984, Emelyanov_2009, Arlot_2014, Saquet_2018, Morgado_2019c}, mutual approximations \citep{Morgado_2016, Morgado_2019a}, radar astrometry \citep{Brozovic_2020}, stellar occultations \citep{Morgado_2019b}, among others.

A stellar occultation occurs when a Solar System object passes in front of a star for an observer. This technique allows the determination of sizes and shapes with kilometer precision and also obtain its position, albedo, density, etc \citep[and references therein]{Sicardy_2011, sora}. It is essential to have good ephemerides of the occulting objects and a good knowledge of the occulted star's position to predict stellar occultations accurately. Thanks to the Gaia Mission \citep{GAIA2016a} and its catalogs -- Gaia DR1, DR2, and EDR3 \citep{GAIA2016b, GAIA2018, GAIA2021} -- the positions of the stars are known at mas-level.

For the Galilean moons, stellar occultations can provide shapes and sizes with uncertainties comparable with those of space probe images, but that is not affected by albedo variations on the satellite surface, which is highly influenced by the solar phase angle. From an astrometric point of view, these events can provide the best ground-based astrometry of these moons, with uncertainties in the order of 1 mas ($\sim$ 3 km at Jupiter distance at opposition). This accuracy is usually one order of magnitude better than other methods \citep{Morgado_2019b}.

Historically, stellar occultations by Galilean satellites has been recorded since 1971 using photoelectric photometers \citep{Hubbard_1972, Carlson_1973}. Nonetheless, few events were observed since and they were organized in the Small Bodies Occultations Database\footnote{Website: \url{https://sbn.psi.edu/pds/resource/occ.html}} using the reduction process described in \cite{Herald_2020}. These events are rare since only stars with magnitude V brighter than 11.5 will provide a magnitude drop higher than 0.5\%.

Between 2019 and 2021, Jupiter was crossing a very dense star region, with the Galactic center as its background \citep{Gomes-Junior_2016}. The probability of a Jovian Moon to occult a bright star increased dramatically during this epoch, allowing the organization of observational campaigns to observe such events. This passage will occur again only in 2030.

Here, we will detail the analysis of three stellar occultations by Galilean moons. These events were: (i) the double chord stellar occultation by Io (JI) on 2021-04-02; (ii) the multi-chord event by Ganymede (JIII) on 2020-12-21; and (iii) the single-chord event by Europa (JII) on 2019-04-06. The first was favorable to Central and South America with positive detections on Aruba and Curacao. The second was favorable to the USA during the twilight. The third was observable from South America, with a positive detection on Peru. 

Also, we re-analyze two previously published stellar occultations by Galilean moons to provide new astrometry in the Gaia EDR3 reference frame. They were the stellar occultation by Europa on 2017-03-31 published by \cite{Morgado_2019b} and the occultation by Ganymede on 2016-04-13 previously analyzed by \cite{Daversa_2017}. 
This project aims at obtaining the astrometric position of the occulting Galilean Moon at the occultation instant for all the listed events. 

This paper is organized as follows. In \Autoref{sec:redu} we described the analysis process, the pipeline and software used. \Autoref{sec:obs} contains the observational details and our results. We give our final remarks in \Autoref{sec:conc}.

\section{Data Analysis}\label{sec:redu}

We used the stacking consecutive images technique and classical photometric pipelines to extract the occultation light curves, as detailed in \Autoref{sec:photometry}. Abrupt opaque edges models, including the effects of diffraction, finite bandwidth, exposure time, and stellar diameter, were fitted to the star dis- and re-appearances behind the satellite at the various stations defined the occultation chords (\Autoref{sec:lightcurve}). As described in \Autoref{sec:astrometry}, we then used the known 3D shape of the objects to find its limb and fitted this limb to the occultation chords. From this fit, we obtained the center of the figure and determine the astrometry of the occulting satellite.

\subsection{Analysis of images and aperture photometry} \label{sec:photometry}

The images obtained in the observational campaigns were saved in different formats, and the first step was converting them to \texttt{.fits} files. From \texttt{.avi} to \texttt{.fits}, we used a proprietary \textsc{Python} software based on \texttt{astropy v4.0.1} \citep{astropy}. When calibration images (bias, dark and flat-field) were available we used them to correct the original images using standard procedures of the Image Reduction and Analysis Facility \citep[\texttt{IRAF,}][]{iraf}.

When observing the Galilean satellites, one should take particular care not to saturate the CCD image, which will quickly happen due to Jupiter's brightness. On the other hand, stellar occultations by bright objects, such as the Galilean moons (visual magnitude about 5.5), will usually have small magnitude drops (mag. $\sim$11 star will have magnitude drop about 1.0\%). So, saturation would usually happen before achieving an adequate Signal to Noise ratio (S/N) to show the magnitude drop. Whenever the magnitude drop was too small to be measured in the single images, we used the stacking consecutive images technique to increase the images S/N at the cost of the time resolution.

Before stacking the images, we measured the target centroid ($x$,~$y$) in the images with a 2D circular symmetric Gaussian fit over pixels within one full-width at half-maximum (FWHM $\propto$ seeing) from the center. For this, we used the Platform for Reduction of Astronomical Images Automatically \citep[\texttt{PRAIA,}][]{Assafin_2011}. The alignment consists of vertical and horizontal shifts for each image ($\Delta x$,~$\Delta y$) relative to a chosen reference image; in our case, the first images of each data-set. After the alignment between the images, the stack was done with proprietary software in \textsc{Python} \citep[Appendix A]{Morgado_2019b}. We manually chose the number of stacked images to get the best compromise between S/N and time resolution.

The next step was the aperture photometry of the target and calibration objects. We used the PRAIA package for this step. Note that during the occultation, the star and the satellite are blended in the same aperture. This combined flux was normalized to unity outside the occultation, using a polynomial fit before and after the event. Finally, nearby satellites were used as photometric calibrators to correct for low-frequency sky transparency fluctuations.

\subsection{Times and projection in the sky plane} \label{sec:lightcurve}

With the normalized light curves, we obtained the immersion (disappearance) and emersion (reappearance) times using the Stellar Occultation Reduction and Analysis \citep[\texttt{SORA,}][]{sora} package. The fitted occultation model considers a sharp-edge occultation model convolved with Fresnel diffraction, stellar diameter (projected at the body distance), CCD bandwidth, and finite integration time.

Using the Jovian ephemeris (here, the \texttt{jup365} and \texttt{de440}), the Gaia EDR3 star positions propagated to the event epoch, and the observer’s position on Earth (latitude, longitude, and height), each dis- or reappearance time is associated with a stellar position relative to the occulting satellite in the sky-plane ($f$, $g$). We expressed this position in kilometers, f (resp. g) being counted positively towards the local east (resp. north) celestial direction. The pair of positions from the same site describes a chord.

\subsection{Limb fit and astrometric result} \label{sec:astrometry}

Each chord extremity is a point at which we can fit the limb of the figure. Here we considered the satellites' 3D size and shape to be known, as observed by many space missions. For instance, (501) Io complex 3D shape was published by \cite{White_2014}. On the other hand, (502) Europa global shape was studied and published by \cite{Nimmo_2007} and (503) Ganymede by \cite{Zubarev_2015}. 

Using the rotational elements of the Galilean satellites (pole coordinates, direction of the prime meridian, and its time variation) as provided by \cite{Archinal_2018}, we calculated their space orientation as seen from Earth at the occultation epoch. In other words, we determined the geocentric sub-observer latitude ($\phi$) and longitude ($\lambda$) at the occultation instant.

After orienting the 3D shape in the sky plane, we obtain its limb and use it to fit the center of the figure, considering the observed chords. This fit was done using a Monte Carlo approach with uniform distributions of initial guesses to test a vast number of simulated central positions ($\sim$1.000.000) for which we compute chi-squared statistics using \Autoref{Eq:ellipse_fit}. 

\begin{equation}
    \chi^2 = \sum_{i=1}^{N}\frac{(r_{i} - r'_{i})^2}{\sigma_{ri}^2}
    \label{Eq:ellipse_fit}
\end{equation}
were $\sigma_{ri}$ stands for the radial component of the  uncertainty in each chord extremity. With this approach we determine the fitted value as the one that minimizes the $\chi^2$, and we determine the marginal 1-$\sigma$ error bar considering the region where $\chi^2 < \chi^2_{min} + 1$, as discussed in \cite{Numerical_Recipes}.

The center positions ($f_c$, $g_c$) can be converted to an astrometric offset between the occulting object' center of figure and the star's position ($\Delta\alpha\cos\delta$, $\Delta\delta$). The Gaia EDR3 catalogs provides the stars' positions at sub-mas-level precision \citep{GAIA2021}, allowing us to calculate astrometric positions of the occulting object with high accuracy. We highlight that the orientation of the coordinate axes is the same as the ICRS, and we set the origin in the geocenter. The uncertainties of the positioning stem from the uncertainties of the fitted center ($\sigma f_{c},~\sigma g_{c}$) and the uncertainties in the star position propagated to the occultation epoch. Note that the resulting uncertainties are usually at mas level, corresponding to a few kilometers at Jupiter's distance, where 1~mas correspond to $\sim$3~km at opposition.

The reduction pipeline do not consider any correction for relativistic deflection by the Sun or other major bodies on the relative position between the occulting satellite and the occulted star.

\section{Observational Campaigns and Results} \label{sec:obs}

The passage of Jupiter on crowded stellar fields created some opportunities to observe stellar occultations of bright stars by the Galilean moons. Between 2019 and 2021, we organized six observational campaigns, three of which resulted in data sets with at least one positive detection. Some of those campaigns involved stars with magnitude fainter than mag 10, and even the stacking consecutive images technique does not allow to retrieve any positive detection.

The observational campaigns were organized in collaboration with citizen astronomers spread in South and North America, including some communities, such as the International Occultation Timing Association (IOTA/USA\footnote{Website: \url{https://occultations.org/}}). We organized the prediction details of the successful campaigns in \Autoref{tb:predictions}. This table shows the occultation date and time, the occulted star' Gaia EDR3 source identifier, its Right Ascension, Declination, and G magnitude. The stars' positions were propagated to the occultation epoch using the rigorous stellar motion described by \cite{Butkevich_2014}. We also corrected from systematic error in the stars' proper motion of Gaia EDR3 as suggested by \cite{Cantat-Gaudin_2021}. We highlight that there are no Gaia flags about duplicity or strange astrometrical behavior for the stars listed here.


\begin{table*}
\begin{center}
\caption{Occulted stars parameters for each observed event as obtained from Gaia EDR3.}
\begin{tabular}{cccccc}
\hline
\hline
Ev. & Date and time UTC & Gaia EDR3          & Right Ascension$^{(*)}$ & Declination$^{(*)}$ & G mag \\
& yyyy-mm-dd hh:mm:ss         & source identifier  &                hh mm ss.sss (mas) & dd mm ss.sss (mas)  &  \\
\hline
\hline

a & 2021-04-02 10:24:00 & 6841731655255386240 & 21$^h$ 43$^m$ 04$^s$.38938 (0.24) & $-$14$^\circ$ 23' 58''.5337 (0.16) & 5.82  \\
b & 2020-12-21 00:49:00 & 6866303987792105856 & 20$^h$ 09$^m$ 33$^s$.49678 (0.12) & $-$20$^\circ$ 35' 38''.7273 (0.08) & 7.38  \\
c & 2019-06-04 02:26:00 & 4114900624661072256 & 17$^h$ 16$^m$ 59$^s$.88385 (0.08) & $-$22$^\circ$ 28' 06''.8486 (0.04) & 9.14  \\
d & 2017-03-31 06:44:00 & 3629528535155010432 & 13$^h$ 12$^m$ 15$^s$.54298 (0.03) & $-$05$^\circ$ 56' 48''.7523 (0.02) & 9.51  \\
e & 2016-04-13 11:57:00 & 3866368596817620352 & 11$^h$ 03$^m$ 41$^s$.27681 (0.02) & $+$07$^\circ$ 34' 54''.6795 (0.02) & 6.99  \\

\hline
\hline
\multicolumn{6}{l}{\rule{0pt}{3.0ex} ${(*)}$ The stars coordinates (RA and Dec.) and their uncertainties were propagated to the occultation epoch with the}\\
\multicolumn{6}{l}{formalism proposed by \cite{Butkevich_2014} using the parameters (proper motion, parallax, radial velocity, etc)}\\
\multicolumn{6}{l}{from Gaia EDR3 \citep{GAIA2021}, corrected as suggested by \cite{Cantat-Gaudin_2021}.}\\

\label{tb:predictions}
\end{tabular}
\end{center}

\end{table*} 

Details about the observational stations that participate in each campaign can be found in Appendix \ref{App:Observation}. As an example, Figure \ref{fig:occ_map} shows the map of the occultation by Ganymede on 2020-12-21. The blue lines represent the size limit of Ganymede. The black dots represent the body's center for a given time, each separated by one minute. The blue dots are the eleven stations that observed this event, here represented by the name of the leading observer. In this map, the shades of gray stand for the night, twilight, and day part of the globe.

\begin{figure}
    \centering
    \includegraphics[width=\linewidth]{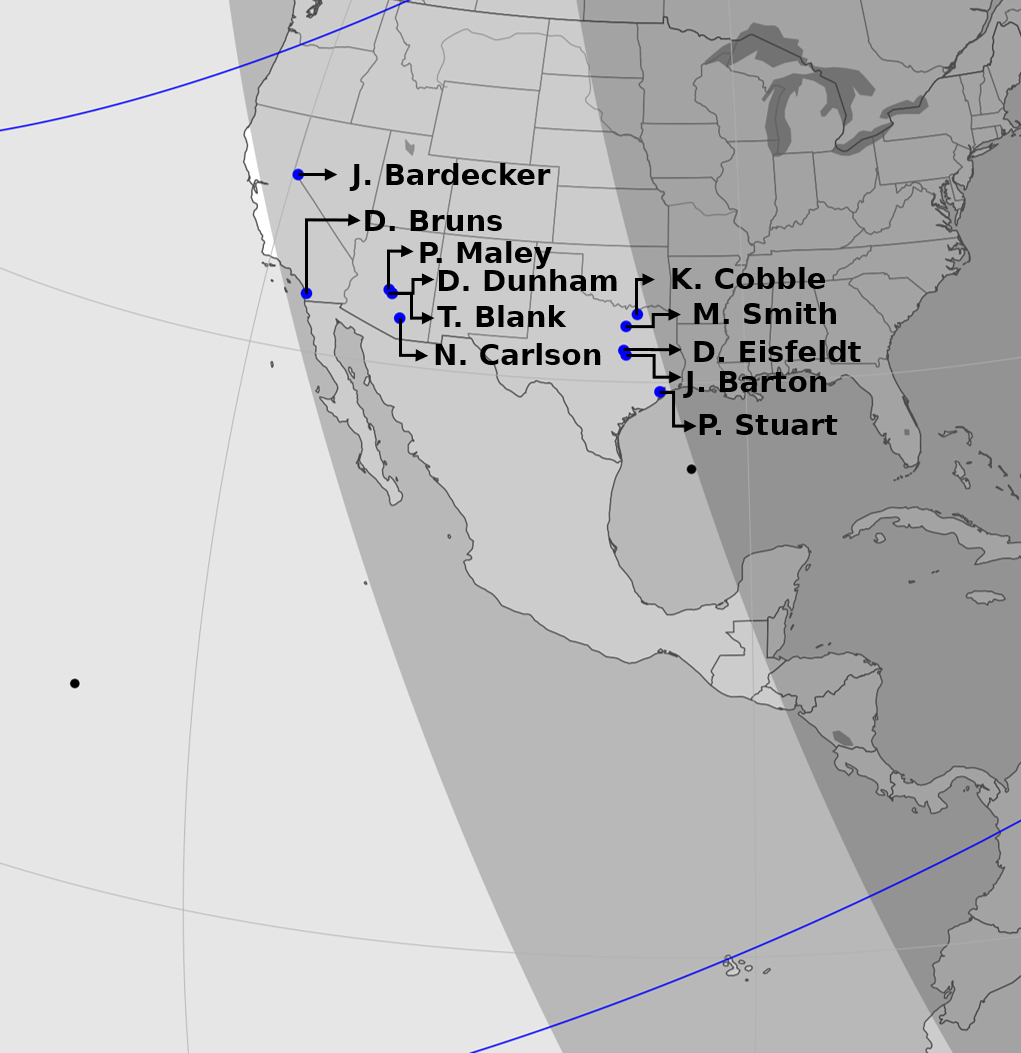}
    \caption{Map of the stellar occultation by Ganymede in 21-12-2020 observed in the United States of America. The blue dots stand for the observational stations that participated in this campaign, each station indicated by its lead observer's name. The blue lines stand for the shadow limits, and the black dots are the center of the shadow, separated by one minute. The shades of gray stand for the night (dark gray), twilight, and day (light gray) part of the globe.}
    \label{fig:occ_map}
\end{figure}

\begin{figure}
    \centering
    \includegraphics[width=\linewidth]{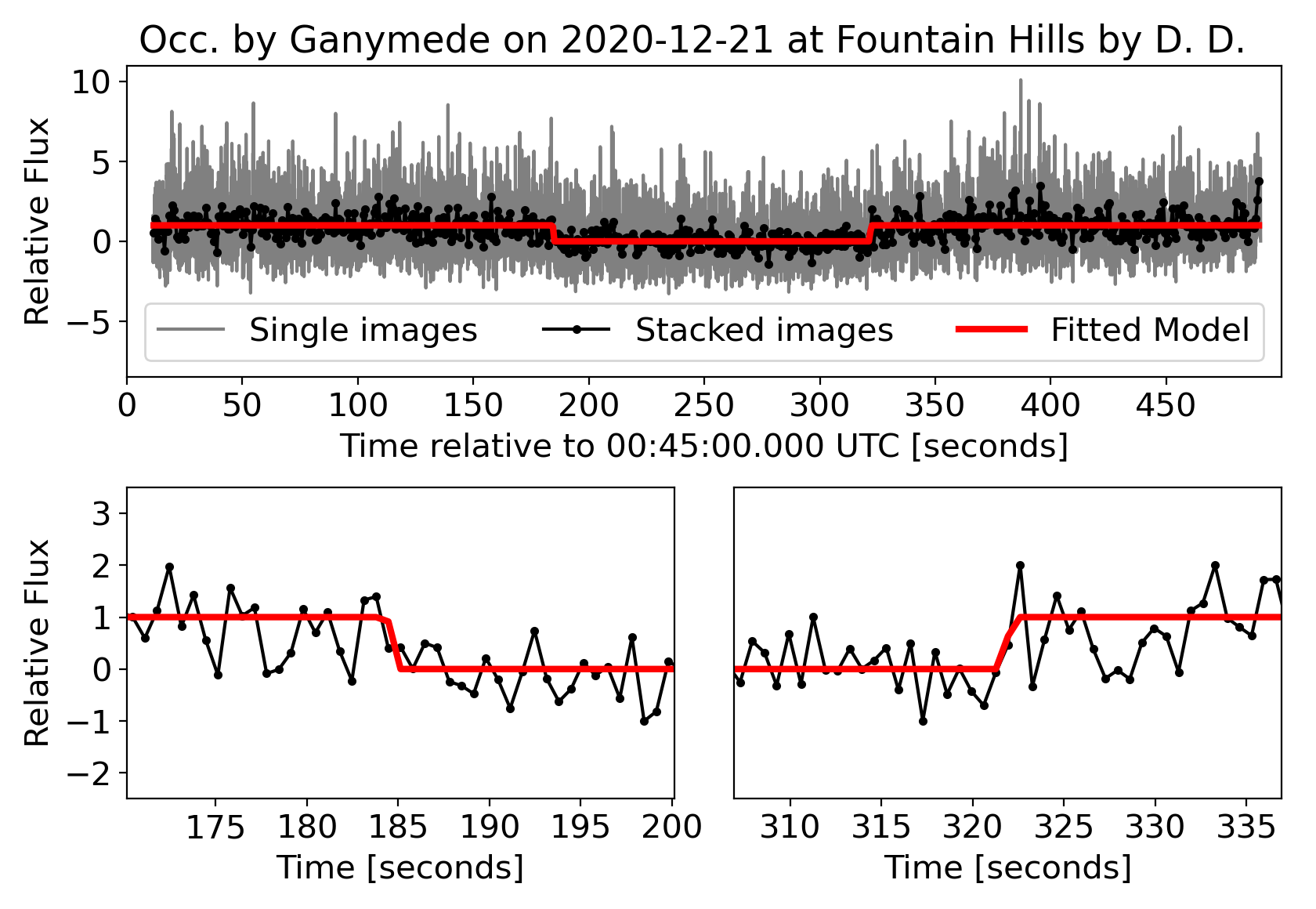}
    \caption{Normalized light curve obtained by D. Dunham of the occultation by Ganymede on 2020-12-21. The gray line shows the photometry of the single images, while the black dots represent the photometry after the stacking of 20 images. The red line represents the fitted model. The bottom panels contain a zoom-in on 30 seconds centered on the immersion and emersion times. We highlight the need for the stacking procedure to detect the occultation in this data set. All the light curves used in this work can be found in Appendix~\ref{App:lightcurves}.}
    \label{fig:lightcurve}
\end{figure}

As mentioned in \Autoref{sec:photometry} all the images in the data sets were analyzed using our pipeline, resulting in light curves like the one shown in \Autoref{fig:lightcurve}. In gray, we see the light flux of every single image, in black the light curve of the stack of 20 consecutive images, and in red the fitted occultation model. Each light curve was analyzed, and the immersion and emersion times were determined as mentioned in \Autoref{sec:lightcurve}. \Autoref{tb:chords} contains the obtained UTC times (immersion and emersion), their uncertainties in seconds, and the minimum chi-squared per degree of freedom ($\chi^2_{pdf}$) obtained for each observational station. All the obtained light curves can be found in \Autoref{App:lightcurves}.

\begin{table*}
\begin{center}
\caption{Fitted times obtained for each light curve with a positive detection.}
\vspace{4pt}
\begin{tabular}{rlcccc} 
\hline 
\hline
\# & Station & Immersion time UTC & Emersion time UTC & Chord length (s) & $\chi^2_{pdf}$  \\ 
\hline
\hline
\multicolumn{6}{c}{\textbf{(a) 2021-04-02 -- Stellar occultation by (501) Io}}\\
\hline
\hline
1 & Savaneta, Aruba -- J. Vrolijk    & 10:19:54.843 (0.420) & 10:20:45.762 (1.278) & \phantom{0}50.919 (1.698) & 0.804  \\
2 & Jan Thiel, Curaçao -- E. Sussenbach & 10:19:50.975 (0.036) & 10:20:56.142 (0.036) & \phantom{0}65.167 (0.072) & 0.755  \\
\hline
\hline
\multicolumn{6}{c}{\textbf{(b) 2020-12-21 -- Stellar occultation by (503) Ganymede}}\\
\hline
\hline
1  & Gardnerville, USA -- J. Bardecker & 00:48:04.694 (0.480) & 00:49:42.632 (0.590) & 103.051 (1.212) & 0.973  \\
2  & San Diego, USA -- D. Bruns        & 00:47:58.782 (0.888) & 00:50:08.367 (1.273) & 129.585 (2.161) & 0.877  \\
3  & Carefree, USA -- P. Maley         & 00:48:04.738 (0.663) & 00:50:20.773 (0.304) & 136.035 (0.967) & 0.864  \\
4  & Fountain Hills, USA -- D. Dunham  & 00:48:04.736 (0.687) & 00:50:21.845 (0.500) & 137.109 (1.188) & 0.820  \\
5  & Fountain Hills, USA -- T. Blank   & 00:48:06.403 (0.751) & 00:50:22.689 (0.513) & 136.286 (1.264) & 1.109  \\
6  & Tucson, USA -- N. Carlson         & $(*)$                & 00:50:28.670 (2.654) & \textgreater 117.670$^{(**)}$   & 0.980  \\
7  & Princeton, USA -- K. Cobble       & 00:48:17.567 (0.778) & 00:50:52.657 (0.242) & 155.090 (1.020) & 0.907  \\
8  & Colleyville, USA -- M. Smith      & 00:48:16.411 (0.774) & 00:50:49.259 (2.548) & 155.257 (1.246) & 0.916  \\
9  & Waco, USA -- D. Eisfeldt          & 00:48:17.664 (1.193) & 00:50:54.283 (1.532) & 152.515 (0.699) & 1.071  \\
10 & Lorena, USA -- J. Barton          & 00:48:18.202 (0.131) & 00:50:54.556 (0.293) & 156.354 (0.424) & 0.922  \\
11 & Clear Lake Shores, USA -- P. Stuart & 00:48:20.539 (0.468) & 00:50:59.476 (0.560) & 158.938 (1.028) & 0.866  \\

\hline
\hline
\multicolumn{6}{c}{\textbf{(c) 2019-06-04 -- Stellar occultation by (502) Europa}}\\
\hline
\hline
1 & CP Cambrune, Peru -- E. Meza       & 02:28:05.033 (0.044) & 02:29:14.311 (0.040) & 069.278 (0.084) & 0.891  \\
\hline
\hline
\multicolumn{6}{c}{\textbf{(d) 2017-03-31 -- Stellar occultation by (502) Europa}}\\
\hline
\hline
1 & OPD -- Brazópolis, Brazil -- M. Assafin    & 06:38:28.550 (0.750) & 06:41:06.150 (0.240) & 157.600 (0.990) & $(***)$  \\
2 & Foz do Iguaçu, Brazil -- D. I. Machado     & 06:39:26.070 (0.770) & 06:41:21.590 (1.070) & 115.520 (1.840) & $(***)$  \\
3 & San Pedro de Atacama, Chile -- A. Maury  & 06:40:38.190 (1.200) & 06:42:22.850 (0.340) & 104.660 (1.540) & $(***)$  \\
\hline
\hline
\multicolumn{6}{c}{\textbf{(e) 2016-04-13 -- Stellar occultation by (503) Ganymede}}\\
\hline
\hline
1 & IRTF -- Mauna Kea, Hawaii -- E. D'Aversa         & 11:51:33.722 (1.218) & 11:56:20.938 (1.239) & 287.216 (2.457) & 0.965  \\
2 & SOAO -- Sobaeksan, South Korea -- T. C. Hinse    & 11:58:02.028 (2.222) & 12:01:51.064 (2.218) & 229.039 (4.440) & 0.579  \\
\hline
\hline
\multicolumn{6}{l}{\rule{0pt}{3.0ex} ${(*)}$ Due to observational issues, the recording started after the immersion time.}\\
\multicolumn{6}{l}{\rule{0pt}{3.0ex} ${(**)}$ The minimum chord length was calculated based on the start of acquisition (00:48:31.334 UTC).}\\
\multicolumn{6}{l}{\rule{0pt}{3.0ex} ${(***)}$ Values from \cite{Morgado_2019b}.}\\

\end{tabular}\label{tb:chords}
\end{center}
\end{table*}

The spectroscopy obtained at the NASA Infrared Telescope Facility (IRTF) using the SpeX\footnote{Website: \url{http://irtfweb.ifa.hawaii.edu/~spex/}} \citep{Rayner_2003} instrument at Mauna Kea needed some additional analysis. This data set consisted of several spectra, each ranging between 0.9 and 2.5 microns, obtained with temporal resolution about 2.8 seconds. As it does not have any additional object in the FoV, it was impossible to calibrate the target's flux for sky fluctuations. We overcome this issue integrating the fluxes within bands Y (between 960 -- 1080 microns), J (1113 -- 1327 microns), H (1476 -- 1784 microns), and K (1995 -- 2385 microns) and obtaining the color light curves considering the flux ratios within bands. This methodology allows the determination of the occultation instants even though the sky transparency was not ideal. That was possible because the sky transparency is attenuated, as it affects the analyzed bands similarly, and the colors of the target star (mag H $-$ mag J = $-0.37$) differ from Ganymede's (mag H $-$ mag J = $+0.10$). \Autoref{app:IRTF} contains the detailed analysis of the IRTF observations.

The 3D shape was then projected to the occultation instant, using the methodology presented in \Autoref{sec:astrometry}. The sub-observer latitude ($\phi$) and longitude ($\lambda$) for each occultation can be found in \Autoref{tb:projection}.

\begin{table}
\begin{center}
\caption{Sub-observer coordinates for the occulting satellite on the event instant.}
\begin{tabular}{ccccc}
\hline
\hline
Ev. & Sat.$^{(*)}$ & Date and time UTC &   $\phi$        & $\lambda$   \\
& & yyyy-mm-dd hh:mm:ss & deg.   & deg.                    \\
\hline
\hline

a & 501 & 2021-04-02 10:24:00 & 196.625 & $+$0.123 \\
b & 503 & 2020-12-21 00:49:00 & 130.072 & $-$1.014 \\
c & 502 & 2019-06-04 02:26:00 & 144.983 & $-$3.057 \\
d & 502 & 2017-03-31 06:44:00 & 270.041 & $-$3.359 \\
e & 503 & 2016-04-13 11:57:00 & 135.783 & $-$1.574 \\

\hline
\hline
\multicolumn{5}{l}{${(*)}$ 501 stands for Io, 502 for Europa and 503 for Ganymede.}\\

\label{tb:projection}
\end{tabular}
\end{center}
\emph{Note}: For (501) Io we used the complex 3D shape as published by \cite{White_2014}. For (502) Europa we used the triaxial shape with axis $a >b >c$ equals to 1562.6, 1560.3 and 1559.5 km as published by \cite{Nimmo_2007}. For (503) Ganymede we used the triaxial ellipsoid with axis 2634.6, 2633.0 and 2631.4 km published by \cite{Zubarev_2015}.

\end{table} 

\begin{figure}
    \centering
    \subfigure[Fit Io2021][Occultation by 501 on 02/04/2021]{\includegraphics[width=.96\linewidth]{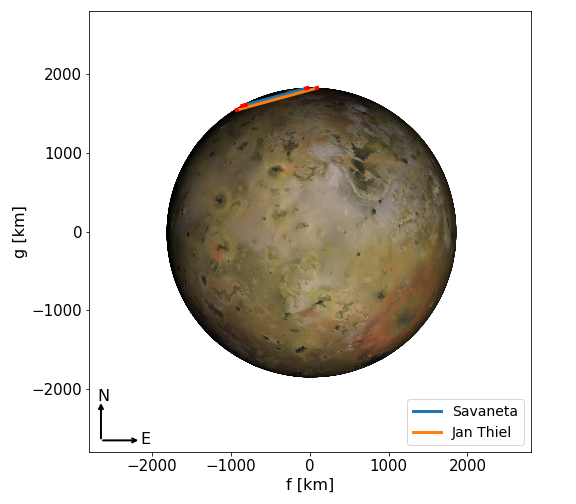} \label{fig:3Dfit_a}
    }
    \subfigure[Fit Ga2020][Occultation by 503 on 21/12/2020]{\includegraphics[width=.96\linewidth]{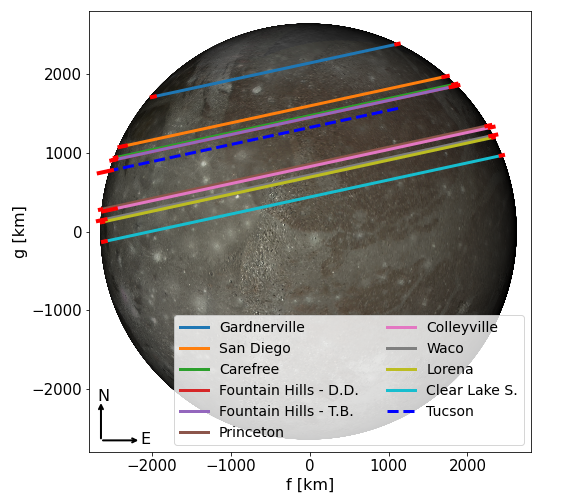} \label{fig:3Dfit_b}
    }
    \caption{We fitted the limb of the Galilean satellites to the stellar occultation chords. In panel (a), we have the stellar occultation by Io on 02/04/2021 observed from South America; (b) by Ganymede on 21/12/2020 observed from North America. The red segment in the extremities stands for the 1$\sigma$ error bar in each dis- and re-appearance time. We call the attention that some redundant chords may be visually indistinguishable.
    \label{fig:3Dfit}}
\end{figure}

The chords for each occultation are then combined, and we fit the limb of the figure as explained in \Autoref{sec:astrometry}. \Autoref{fig:3Dfit} contains the chords and the fitted 3D limb, where each panel stands for an occultation. The obtained astrometric positions were organized in \Autoref{tb:positions}.

Our results consist of geocentric astrometric positions with uncertainties in the mas level. \Autoref{tb:offsets} contains the differences between the observed positions and different geocentric ephemeris.

The 3D models used here for Europa and Ganymede are based on triaxial ellipsoids representing the global shapes of these satellites, meaning that topographic features were not taken into consideration. Nonetheless, based on space-mission data, the expected topographic features of these satellites ranges below 1 km for Europa and 2 km for Ganymede \citep{Thomas_1997, Nimmo_2007, Zubarev_2015}. Furthermore, those values are smaller than the uncertainties obtained in the chords extremities.

All the observations for the campaigns between 2019 and 2021 went through our entire reduction pipeline. On the other hand, the re-analysis of the occultation observed in 2017-03-31 and published by \cite{Morgado_2019b} started with the times that were already published, so no photometry and time determination was done in the context of this project. For the occultation observed in 2016-04-13, the photometric analysis was done previously by \cite{Daversa_2017}, the light curves were then included in our pipeline, and the dis- and re-appearance times were determined. 

\begin{figure*}
    \centering
    \setcounter{subfigure}{4}
    \subfigure[Fit Eu2019][Occultation by 502 on 04/06/2019]{\includegraphics[width=.48\linewidth]{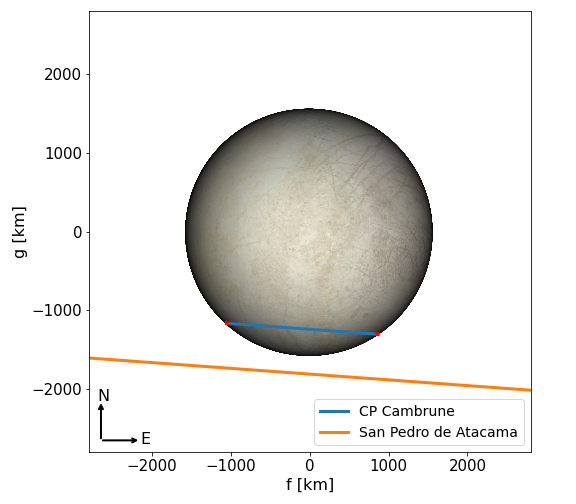} \label{fig:3Dfit_c}
    }
    \subfigure[Fit Eu2017][Occultation by 502 on 31/03/2017]{\includegraphics[width=.48\linewidth]{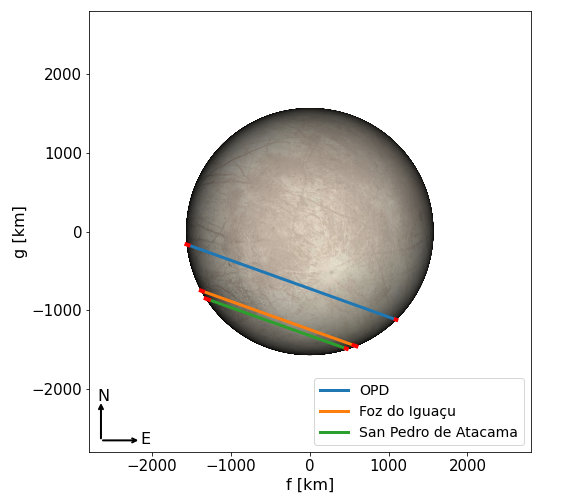} \label{fig:3Dfit_d}
    }
    \subfigure[Fit Ga2016][Occultation by 503 on 2016-04-13]{\includegraphics[width=0.48\linewidth]{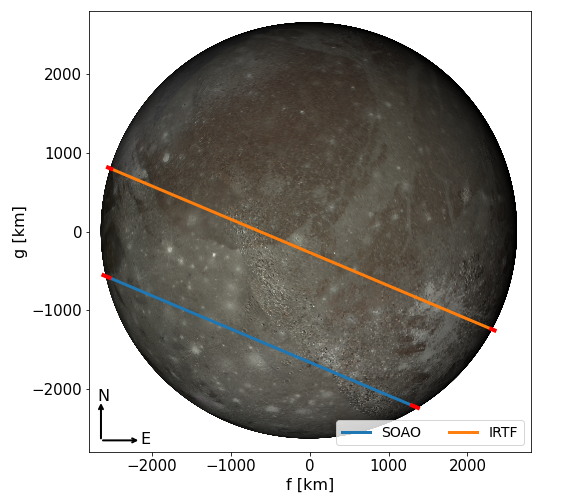} \label{fig:3Dfit_e}
    }
    \caption{Same as Figure \ref{fig:3Dfit} for events: (c) by Europa on 04/06/2019 observed from South America; (d) by Europa on 31/03/2017 observed from South America and previously published by \cite{Morgado_2019b}. (e) by Ganymede on 13/04/2016 observed from North America and previously published by \cite{Daversa_2017}.
\label{fig:3Dfit_2}}
\end{figure*}

\begin{table*}
\begin{center}
\caption{Astrometric positions for each event.}
\begin{tabular}{ccccc}
\hline
\hline
Event & Satellite$^{(*)}$ & Date and time UTC &  Right Ascension$^{(**)}$ & Declination$^{(**)}$ \\
    &              & yyyy-mm-dd hh:mm:ss & hh mm ss.sss (mas) & dd mm ss.sss (mas) \\

\hline
\hline

a & 501 & 2021-04-02 10:24:00.000  & 21$^h$ 43$^m$ 04$^s$.37583 (1.1) & $-$14$^\circ$ 23' 58''.1536 (0.7) \\
b & 503 & 2020-12-21 00:49:00.000  & 20$^h$ 09$^m$ 33$^s$.56022 (0.9) & $-$20$^\circ$ 35' 38''.0137 (1.7) \\
c & 502 & 2019-06-04 02:26:00.000  & 17$^h$ 16$^m$ 59$^s$.89400 (1.1) & $-$22$^\circ$ 28' 06''.5375 (1.1) \\
d & 502 & 2017-03-31 06:44:00.000  & 13$^h$ 12$^m$ 15$^s$.54781 (1.9) & $-$05$^\circ$ 56' 48''.6987 (1.6) \\
e & 503 & 2016-04-13 11:57:00.000  & 11$^h$ 03$^m$ 41$^s$.32089 (4.1) & $+$07$^\circ$ 34' 55''.6614 (4.7) \\

\hline
\hline
\multicolumn{5}{l}{\rule{0pt}{3.0ex} $(*)$ 501 stands for Io, 502 for Europa and 503 for Ganymede}\\
\multicolumn{5}{l}{\rule{0pt}{3.0ex} $(**)$ These positions assume the EDR3 star positions given in Table~\ref{tb:predictions}}\\

\label{tb:positions}
\end{tabular}
\end{center}

\end{table*}

\begin{table*}
\begin{center}
\caption{Offsets in mas between the obtained positions for each event and different ephemerides.}
\begin{tabular}{ccccccccccc}
\hline
\hline
Event & \multicolumn{2}{c}{Uncertainties (mas)}& \multicolumn{2}{c}{$O - C_1$ (mas)} & \multicolumn{2}{c}{$O - C_2$ (mas)} &  \multicolumn{2}{c}{$O - C_3$ (mas)} & \multicolumn{2}{c}{$O - C_4$ (mas)} \\
    & $\Delta RA$   & $\Delta Dec$  & $\Delta RA$   & $\Delta Dec$  &  $\Delta RA$   & $\Delta Dec$  &  $\Delta RA$   & $\Delta Dec$ &  $\Delta RA$   & $\Delta Dec$ \\

\hline
\hline

a & 1.1 & 0.7 & $+$05.5 & $-$02.9 & $-$01.3 & $-$01.5 & $+$06.7 & $-$05.1 & $-$00.1 & $-$03.7 \\
b & 0.9 & 1.7 & $-$04.2 & $-$00.1 & $-$03.4 & $-$01.6 & $-$03.2 & $-$02.4 & $-$02.4 & $-$04.0 \\
c & 1.1 & 1.1 & $-$04.1 & $-$03.1 & $-$08.0 & $-$16.8 & $-$03.9 & $-$05.8 & $-$07.8 & $-$19.5 \\
d & 1.9 & 1.6 & $-$00.2 & $-$00.3 & $-$06.3 & $-$10.6 & $+$00.5 & $+$00.3 & $-$05.7 & $-$10.0 \\
e & 4.1 & 4.7 & $-$02.3 & $+$07.5 & $-$02.2 & $+$09.5 & $-$01.1 & $+$09.4 & $-$01.0 & $+$11.4 \\
\hline

\hline
\hline
\multicolumn{11}{l}{\rule{0pt}{3.0ex} \emph{Note}: $C_1$ is using ephemeris \texttt{de440} and \texttt{jup365}; $C_2$ is \texttt{de440} and NOE-5-2021; $C_3$ is ephemeris}\\
\multicolumn{11}{l}{INPOP19 and \texttt{jup365}; $C_4$ is INPOP19 and NOE-5-2021.}\\

\label{tb:offsets}
\end{tabular}
\end{center}
\end{table*}

\section{Final remarks} \label{sec:conc}

Between 2019 and 2021, we organized and observed stellar occultation campaigns by Galilean moons. Citizen-astronomers did many of the observations reported here with telescopes of small apertures. These campaigns aimed to obtain positions with uncertainties in the mas level.

Stacking consecutive images techniques were used to improve the Signal to Noise ratio, thus emphasizing the low magnitude drop of the occultation. The known 3D shape and size of the Galilean satellites were used here. They were oriented using the sub-observer angles and the formalism presented by \cite{Archinal_2018}. 

From the six organized campaigns, three had at least one positive detection. We also revisited two other stellar occultations previously published by \cite{Morgado_2019b} and \cite{Daversa_2017}, now using the Gaia EDR3 stellar catalog. This project resulted in five positions with a typical uncertainty below 2 mas ($\sim 6$ km at Jupiter distance). These precise positions also allow us to identify significant offsets in the ephemerides. 

The obtained positions reported here can be used to reconsider the accuracy of the ephemerides (for both Jupiter and its moons) that are currently used to prepare JUICE and Europa Clipper space missions. Moreover, it has been demonstrated that both missions will face an almost ill-posed problem during radio-science data inversion because of their lack of Io's flybys \citep{Dirkx_2017}. The present paper demonstrates the huge potential of stellar occultations observed from the ground to mitigate the former issue. Another key aspect of stellar occultations is to provide information on the center of figure of the moons, while radio-science data are sensitive to their center of mass. Such shift is unknown for most Solar system objects but could be quantified by merging radio-science and stellar occultation data. Moreover, that should motivate observers worldwide to invest time and energy in observing these occultations by the Galilean satellites. Last but not least, the current data can evidently be used to improve the current estimations of tidal dissipation within Jupiter and Io \citep{Lainey_2009}.

\clearpage
\section*{acknowledgments}
This work was carried out within the “Lucky Star" umbrella that agglomerates the efforts of the Paris, Granada and Rio teams, which is funded by the European Research Council under the European Community’s H2020 (ERC Grant Agreement No. 669416). This research made use of \textsc{sora}, a python package for stellar occultations reduction and analysis, developed with the support of ERC Lucky Star and LIneA/Brazil, within the collaboration of Rio-Paris-Granada teams. This work has made use of data from the European Space Agency (ESA) mission Gaia (\url{https://www.cosmos.esa.int/gaia}), processed by the Gaia Data Processing and Analysis Consortium (DPAC, \url{https://www.cosmos.esa.int/web/gaia/dpac/consortium}).
Part of this research is suported by INCT do e-Universo, Brazil (CNPQ grants 465376/2014-2).
This work was supported by CNES, focused on Juice.
Based in part on observations made at the Laborat\'orio Nacional de Astrof\'isica (LNA), Itajub\'a-MG, Brazil.
The following authors acknowledge the respective 
i) CNPq grants: 
BEM 150612/2020-6; 
FB-R 314772/2020-0; 
RV-M 304544/2017-5, 401903/2016-8; 
MA 427700/2018-3, 310683/2017-3, 473002/2013-2; 
RS and OCW 305210/2018-1.
ii) CAPES/Cofecub grant:
BEM 394/2016-05.
ii) FAPERJ grants: 
MA E-26/111.488/2013. 
iii) FAPESP grants: 
ARGJr 2018/11239-8; 
RS and OCW 2016/24561-0.
Italian coauthors thank ASI and INAF for the financial support through the "Accordo ASI-INAF n. 2018-25-HH.0". The Infrared Telescope Facility is operated by the University of Hawaii under contract NNH14CK55B with the National Aeronautics and Space Administration. We express special thanks to Bobby Bus as support astronomer for IRTF observations. The Sobaeksan Observatory facility SOAO is managed by the Korean Astronomy and Space Science Istitute (KASI).
A portion of this research was carried out at the Jet Propulsion Laboratory, California Institute of Technology, under a contract with the National Aeronautics and Space Administration (80NM0018D0004).
TCH and YK would like to thank the staff at the SOAO observatory for fruitful discussions on astronomical data presented in this paper.
RS acknowledges support by the DFG German Research Foundation project 446102036.
RS and OCW thanks the Brazilian Federal Agency for Support and Evaluation of Graduate Education (CAPES), in the scope of the Program CAPES-PrInt, process number 88887.310463/2018-00, International Cooperation Project number 3266.
This research received financial support from the National Research Foundation (NRF; No. 2019R1I1A1A01059609). 
This study was financed in part by the Coordenação de Aperfeiçoamento de Pessoal de Nível Superior - Brasil (CAPES) - Finance Code 001.


\appendix

\section{Observational Circumstance} \label{App:Observation}

Table \ref{tb:obs_sites_1} summarize the observational circumstances of each station for the four stellar occultations presented here (name of station, coordinates, observers, telescope aperture, detector, band, exposure and  cycle times, and the light curve Root Mean Square (RMS) noise. The status of each observation is mentioned (positive or negative detection). Overcast weather and instrumental issues are also indicated. The information relative to the occultation by Europa in 2017, March $31^{st}$, was not added in this table as it can be found in \cite{Morgado_2019b}.

\begin{table*}
\begin{center}
\caption{Observational stations, technical details, and circumstance for the stellar occultations on this project.}
\begin{tabular}{llccccc}
\hline
\hline
\textbf{Status}         & \textbf{Longitude} & \textbf{Observers} & \textbf{Telescope Aperture} & \textbf{Single cycle time (s)}    & \textbf{Light-curve} \\
\textbf{Station}           & \textbf{Latitude}  &                    & \textbf{CCD}       & \textbf{Stacked cycle time (s)} & \textbf{rms} \\
\textbf{Country}        & \textbf{Altitude}  &                    & \textbf{Timing}   &       & \\
\hline
\hline

\vspace{0.002cm}\\
\multicolumn{6}{c}{\textbf{(a) 2021-04-02 -- Stellar occultation by (501) Io}}\\
\hline
\hline
Positive             & \phantom{0}69$^o$ 57' 04.0" W            & J. Vrolijk   & 20.3 cm         & 0.026 & 0.504 \\
Savaneta             &  \phantom{0}12$^o$ 27' 10.8" N &              & ZWO ASI224MC    & -- \\
Aruba                & 17 m                         &              & NTP \\ 

\hline
Positive             & \phantom{0}68$^o$ 52' 23.0" W & E. Sussenbach & 27.9 cm         & 0.095 & 0.086 \\
Jan Thiel            & \phantom{0}12$^o$ 05' 37.0" N &          & ZWO ASI462MC & -- \\
Curaçao              & 18 m &       & NTP \\ 
\hline
Weather overcast     & \phantom{0}70$^o$ 40' 42.5" W  & E. Meza    & 100.0 cm      & -- & -- \\
CP Cambrune               & \phantom{0}16$^o$ 49' 41.3" S  &            & ZWO ASI178MM  & -- \\
Peru                 & 3305 m                         &            & GPS \\ 
\hline
Daylight             & \phantom{0}49$^o$ 20' 23.7" W  & F. Braga-Ribas    & 20.3 cm      & -- & -- \\
Curitiba             & \phantom{0}25$^o$ 26' 10.4" S  &                   & QHY174-GPS  & -- \\
Brazil               & 935 m                         &                    & GPS \\

\hline
\hline
\vspace{0.002cm}\\
\multicolumn{6}{c}{\textbf{(b) 2020-12-21 -- Stellar occultation by (503) Ganymede}}\\
\hline
\hline
Positive             & 119$^o$ 40' 20.3" W & J. Bardecker & 30.5 cm         & 0.033 & 0.905 \\
Gardnerville         &  \phantom{0}38$^o$ 53' 23.5" N &          & Watec 910HX  & 0.667 \\
USA                  & 1524 m &       & IOTA-VTI \\ 

\hline
Positive             & 117$^o$ 09' 37.0" W & D. Bruns & 27.9 cm         & 0.100 & 0.680 \\
San Diego            &  \phantom{0}32$^o$ 56' 16.7" N &          & ZWO ASI1600MM & 1.000 \\
USA                  & 81 m &       & SNTP \\ 

\hline
Positive             & 111$^o$ 57' 08.0" W& P. Maley & 20.0 cm         & 0.033 & 0.874 \\
Carefree             &  \phantom{0}33$^o$ 48' 42.9" N&          & Night Eagle Astro & 0.667 \\
USA                  & 654 m &       & IOTA-VTI \\ 

\hline
Positive             & 111$^o$ 43' 39.0" W& D. Dunham & 12.7 cm         & 0.033 & 0.727\\
Fountain Hills       &  \phantom{0}33$^o$ 37' 28.3" N& J. Dunham & Night Eagle Astro  & 0.667 \\
USA                  & 520 m &       & IOTA-VTI \\ 

\hline
Positive             & 111$^o$ 43' 35.3" W& T. Blank  & 35.6 cm         & 0.033 & 0.931 \\
Fountain Hills       &  \phantom{0}33$^o$ 37' 21.2" N&            & Watec 910HX & 0.667 \\
USA                  & 515 m &            & IOTA-VTI \\ 

\hline
Positive             & 111$^o$ 01' 48.5" W& N. Carlson & 23.5 cm         & 0.040 & 1.126 \\
Tucson               &  \phantom{0}32$^o$ 25' 00.0" N&          & Night Eagle Astro  & 0.800 \\
USA                  & 825 m &       & Visual \\ 

\hline
Positive             & \phantom{0}96$^o$ 29' 19.6" W& K. Cobble & 12.7 cm         & 0.034 & 0.628 \\
Princeton            & \phantom{0}33$^o$ 29' 30.7" N&          & QHY 174 GPS & 0.669 \\
USA                  & 175 m &       & GPS \\ 

\hline
Positive             & \phantom{0}97$^o$ 08' 20.8" W& M. Smith & 12.7 cm         & 0.033 & 0.857 \\
Colleyville          & \phantom{0}32$^o$ 52' 50.2" N&          & Watec 902H  & 0.667\\
USA                  & 160 m &       & IOTA-VTI \\ 

\hline
\hline
\label{tb:obs_sites_1}
\end{tabular}
\end{center}

\end{table*}

\setcounter{table}{5}

\begin{table*}
\begin{center}
\caption{\textbf{[Cont.]} Observational stations, technical details, and circumstance for the stellar occultations on this project.}
\begin{tabular}{llccccc}
\hline
\hline
\textbf{Status}         & \textbf{Longitude} & \textbf{Observers} & \textbf{Telescope Aperture} & \textbf{Single cycle time (s)}    & \textbf{Light-curve} \\
\textbf{Station}           & \textbf{Latitude}  &                    & \textbf{CCD}       & \textbf{Stacked cycle time (s)} & \textbf{rms} \\
\textbf{Country}        & \textbf{Altitude}  &                    & \textbf{Timing}   &       & \\
\hline
\hline
\vspace{0.002cm}\\
\multicolumn{6}{c}{\textbf{[Cont.] (b) 2020-12-21 -- Stellar occultation by (503) Ganymede}}\\
\hline
\hline
Positive             & \phantom{0}97$^o$ 14' 59.8" W& D. Eisfeldt & 20.3 cm         & 0.033 & 0.893 \\
Waco                 & \phantom{0}31$^o$ 37' 44.6" N&          & Night Eagle Astro  & 0.667 \\
USA                  & 161 m &       & IOTA-VTI \\ 

\hline
Positive             & \phantom{0}97$^o$ 05' 04.6" W& J. Barton & 31.7 cm         & 0.033 & 0.656 \\
Lorena               & \phantom{0}31$^o$ 24' 47.9" N&          & Watec 902H   & 0.667\\
USA                  & 134 m &       & IOTA-VTI \\ 

\hline
Positive             & \phantom{0}95$^o$ 02' 06.6" W& P. Stuart & 20.3 cm         & 0.034 & 0.650 \\
Clear Lake Shores    & \phantom{0}29$^o$ 32' 55.5" N&          & Watec 910BD & 0.718 \\
USA                  & 16 m &       & IOTA-VTI \\ 
\hline
\hline
\vspace{0.002cm}\\
\multicolumn{6}{c}{\textbf{(c) 2019-06-04 -- Stellar occultation by (502) Europa}}\\
\hline
\hline
Positive             & \phantom{0}70$^o$ 40' 42.5" W  & E. Meza    & 212.0 cm      & 0.075 & 0.587 \\
CP Cambrune          & \phantom{0}16$^o$ 49' 41.3" S  &            & ZWO ASI178MM  & -- \\
Peru                 & 3305 m                         &            & GPS \\ 

\hline
Negative             & \phantom{0}68$^o$ 10' 48.0" W  & A. Maury   & 15.5 cm       & 3.570 & 0.610 \\
San Pedro de Atacama & \phantom{0}22$^o$ 57' 08.0" S  &            & STX-16803     & -- \\
Chile                & 2397 m                         &            & GPS \\ 

\hline
No detection         & \phantom{0}43$^o$ 59' 03.1" W  & C. Jacques & 45.0 cm       & 1.690 & 2.169 \\
Oliveira             & \phantom{0}19$^o$ 52' 55.0" S  &            & ML FLI16803     & -- \\
Brazil               & 982 m                          &            & GPS \\ 

\hline
Target saturated     & \phantom{0}40$^o$ 19' 00.0" W  & M. Malacarne & 35.0 cm           & 8.000 & -- \\
Vitoria              & \phantom{0}20$^o$ 17' 52.0" S  &              & SBIG/ST-8X-ME     & -- \\
Brazil               & 26 m                          &               & NTP \\ 

\hline
Weather overcast     & \phantom{0}50$^o$ 05' 56.6" W  & C. L. Perreira & 40.0 cm       & -- & -- \\
Ponta Grossa         & \phantom{0}25$^o$ 05' 22.5" S  & M. Emilio      & Merlin/Raptor & -- \\
Brazil               & 910 m                          &                & GPS \\ 

\hline
Weather overcast     & \phantom{0}45$^o$ 34' 57.5" W & B. Morgado         & 60 cm          & -- & -- \\
OPD -- Brazópolis    & \phantom{0}22$^o$ 32' 07.8" S &                    & Andor/Ixon-EM  & -- & \\
Brazil               & 1864 m                        &                    & GPS &  \\ 

\hline
Weather overcast     & \phantom{0}45$^o$ 11' 25.0" W & R. Sfair           & 40 cm          & -- & -- \\
Guaratinguetá        & \phantom{0}22$^o$ 48' 34.0" S & A. R. Gomes-Júnior & Merlin/Raptor  & -- & \\
Brazil               & 567 m                         & O. C. Winter       & GPS &  \\ 
\hline
\hline
\vspace{0.002cm}\\
\multicolumn{6}{c}{\textbf{(e) 2016-04-13 -- Stellar occultation by (503) Ganymede}}\\
\hline
\hline
Positive             & 155$^o$ 28' 18.0" W            & E. D'Aversa   & 320.0 cm        & 2.817 & 0.145 \\
IRTF -- Mauna Kea    &  \phantom{0}19$^o$ 49' 34.0" N & T. Oliva & SpeX            & -- \\
Hawaii               & 4150 m                         & G. Sindoni           & GPS \\ 

\hline
Positive             & 128$^o$ 27' 27.6" E            & T. C. Hinse   & 61.0 cm         & 6.048 & 0.066 \\
SOAO -- Sobaeksan    &  \phantom{0}36$^o$ 56' 03.9" N & Y. Kim        & ProLine PL16803      & -- \\
South Korea          & 1378 m                         &               & NTP \\ 

\hline
\hline
\label{tb:obs_sites_2}
\end{tabular}
\end{center}

\end{table*}

\section{Light Curves}\label{App:lightcurves}

This section provides the plots of the normalized light curves vs. UTC analyzed in this work. There are 17 light curves obtained during four occultations observed between 2016 and 2021. Here we are excluding the light curves from the 2017 March 31 that were published in \cite{Morgado_2019b}. The events are listed in inverse chronological order, and the light curves are plotted from the northernmost to the southernmost stations. The date, event, and observer are indicated in the title and label of each figure. The upper panel contains the complete normalized light curve (black dots) and the fitted model (red line). The bottom panels have expanded views of a few seconds centered on the star's immersion and emersion behind the main body. We also show in the upper panel the light curve before the stacking consecutive images technique (gray line) whenever it was used. The normalization was done so the flux outside the event would be one and the flux within the event to be zero, we highlight that the target never disappears from the images.

\begin{figure}[h]
\centering
\includegraphics[width=0.49\textwidth]{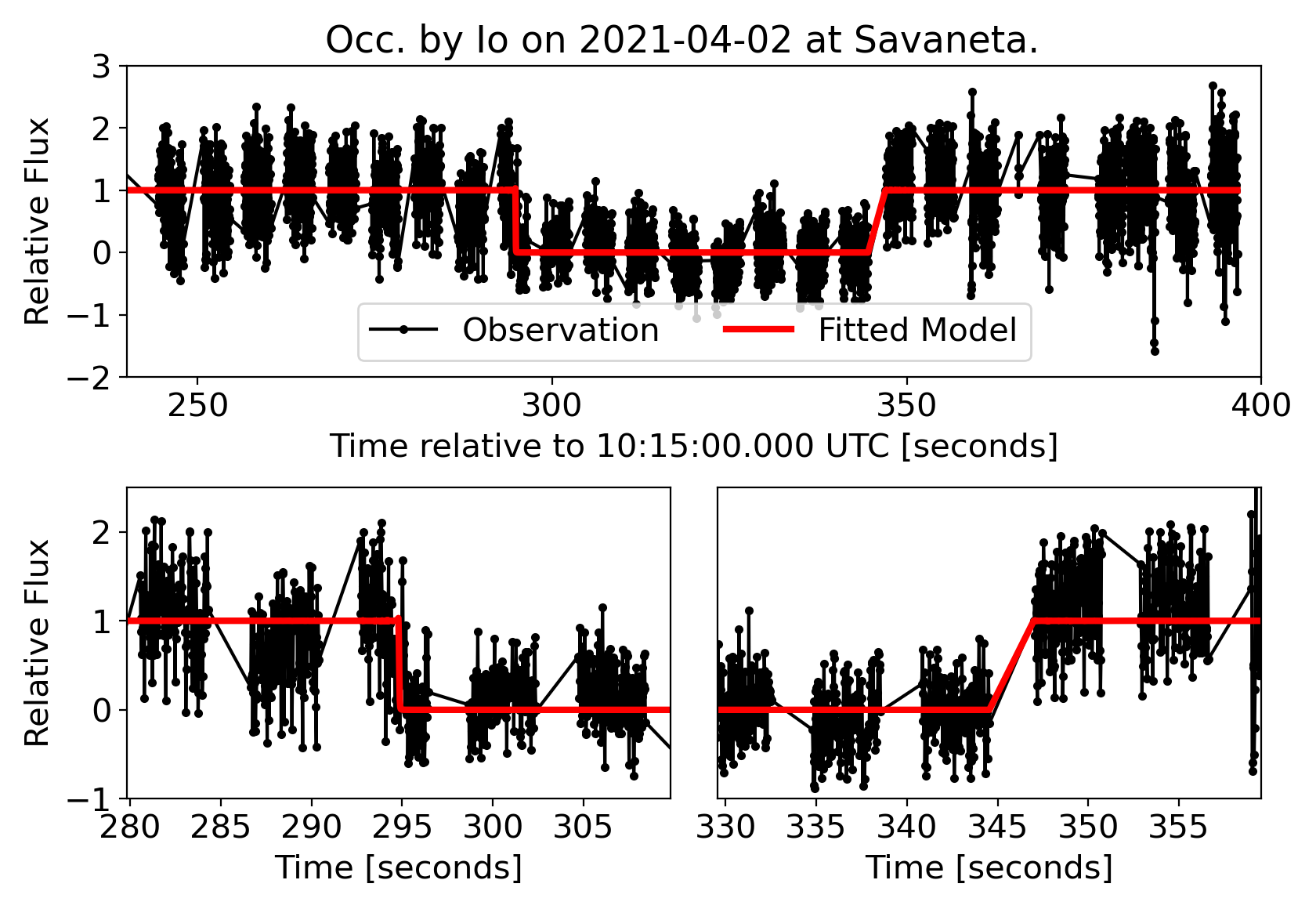}
\caption{Light curve obtained in Savaneta on 2021-04-02 by J. Vrolijk. The occulting satellite, date and observational station are indicated in the title and label. The upper panel contains the complete normalized light curve (black dots) and the fitted model (red line). The bottom panels contain zoom-in views of a few seconds each centered on the immersions and emersions of the star behind the main occulting satellite.}
\end{figure}               

\begin{figure}[h]
\centering
\includegraphics[width=0.49\textwidth]{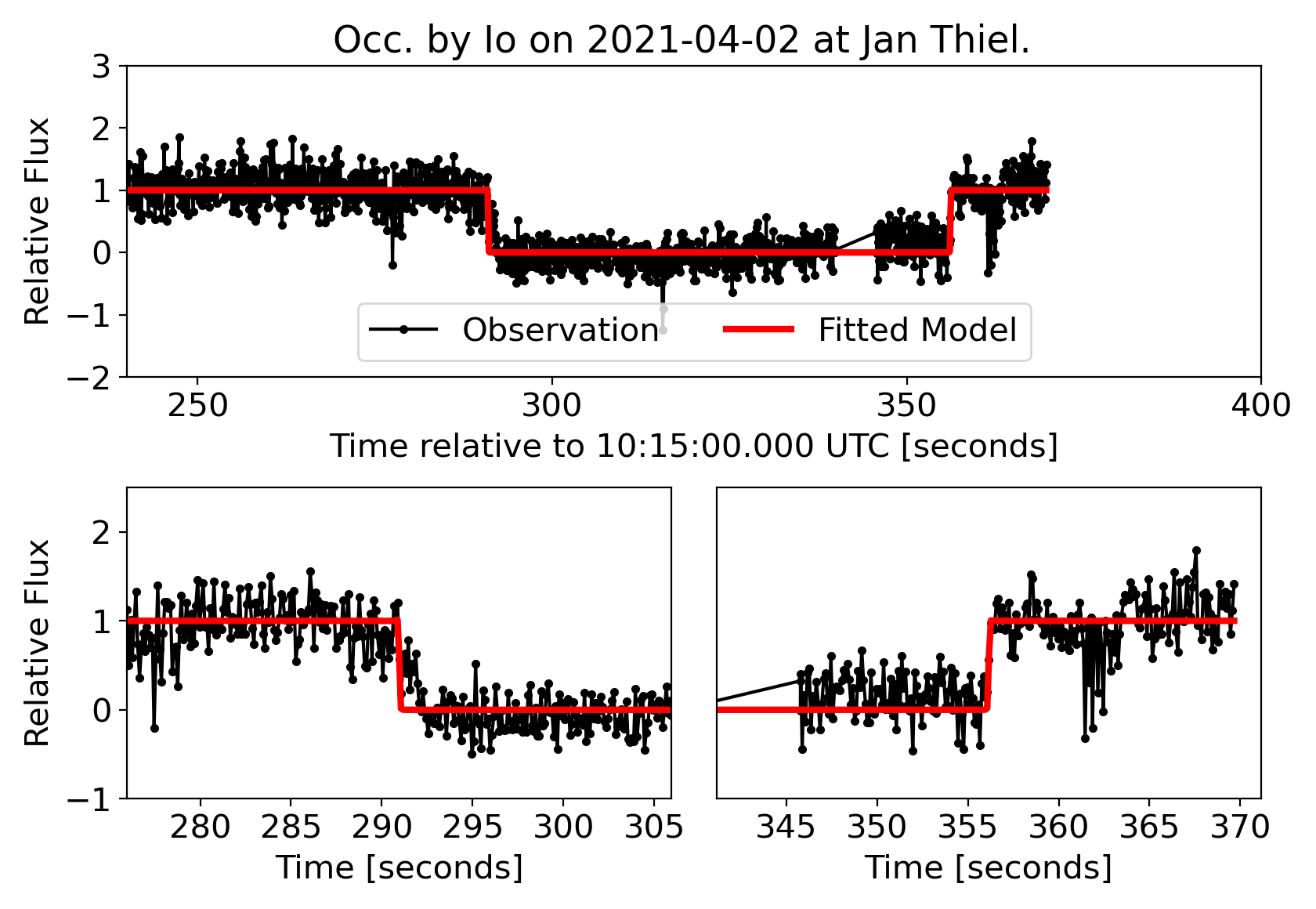}
\caption{Occ. of Io on 2021-04-02 at Jan Thiel by E. Sussenbach.}
\end{figure}

\begin{figure}[h]
\centering
\includegraphics[width=0.49\textwidth]{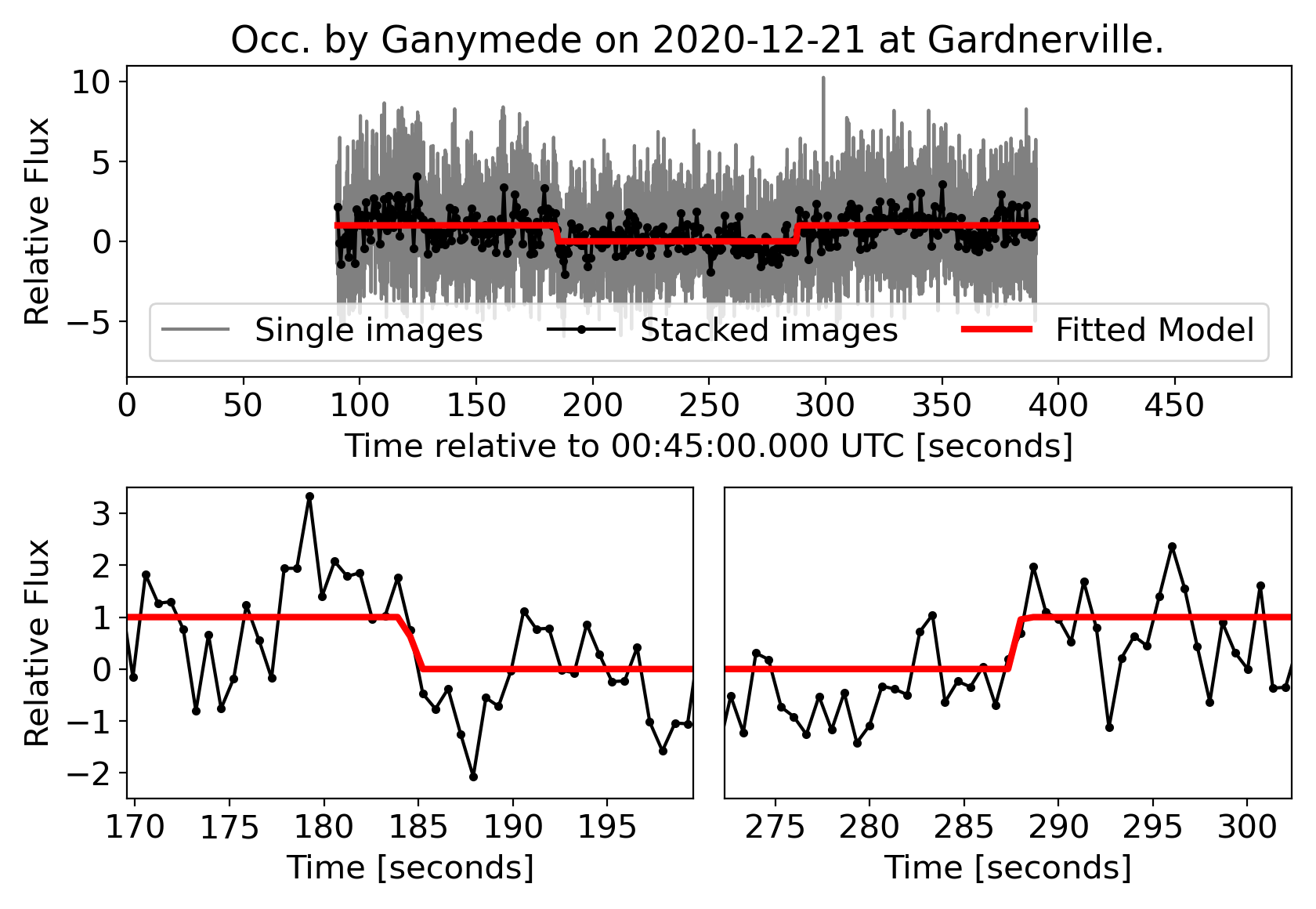}
\caption{Occ. of Ganymede on 2020-12-21 at Gardnerville by J. Bardecker.}

\end{figure}               

\begin{figure}[h]
\centering
\includegraphics[width=0.49\textwidth]{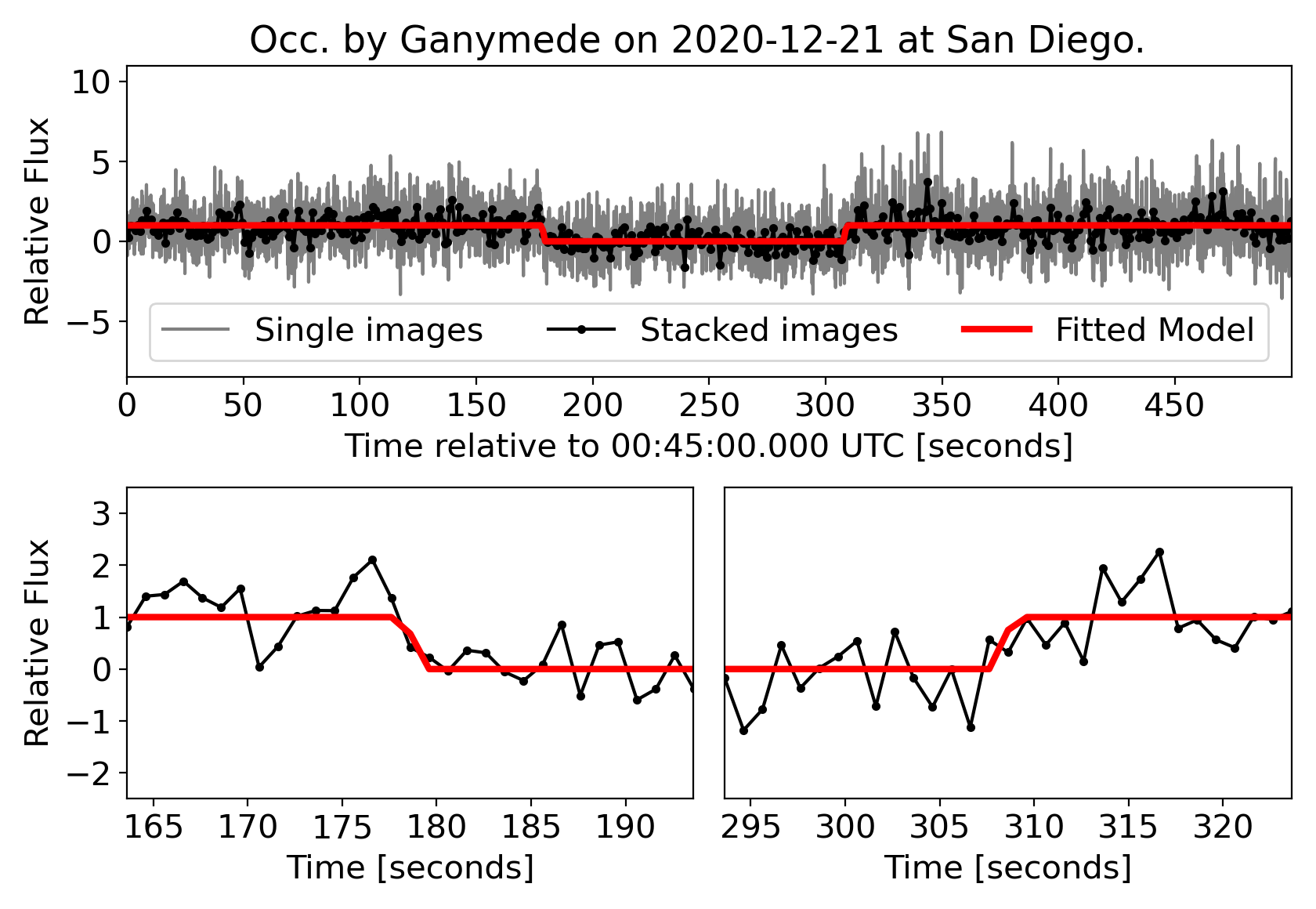}
\caption{Occ. of Ganymede on 2020-12-21 at San Diego by D. Bruns.}
\end{figure}               

\begin{figure}[h]
\centering
\includegraphics[width=0.49\textwidth]{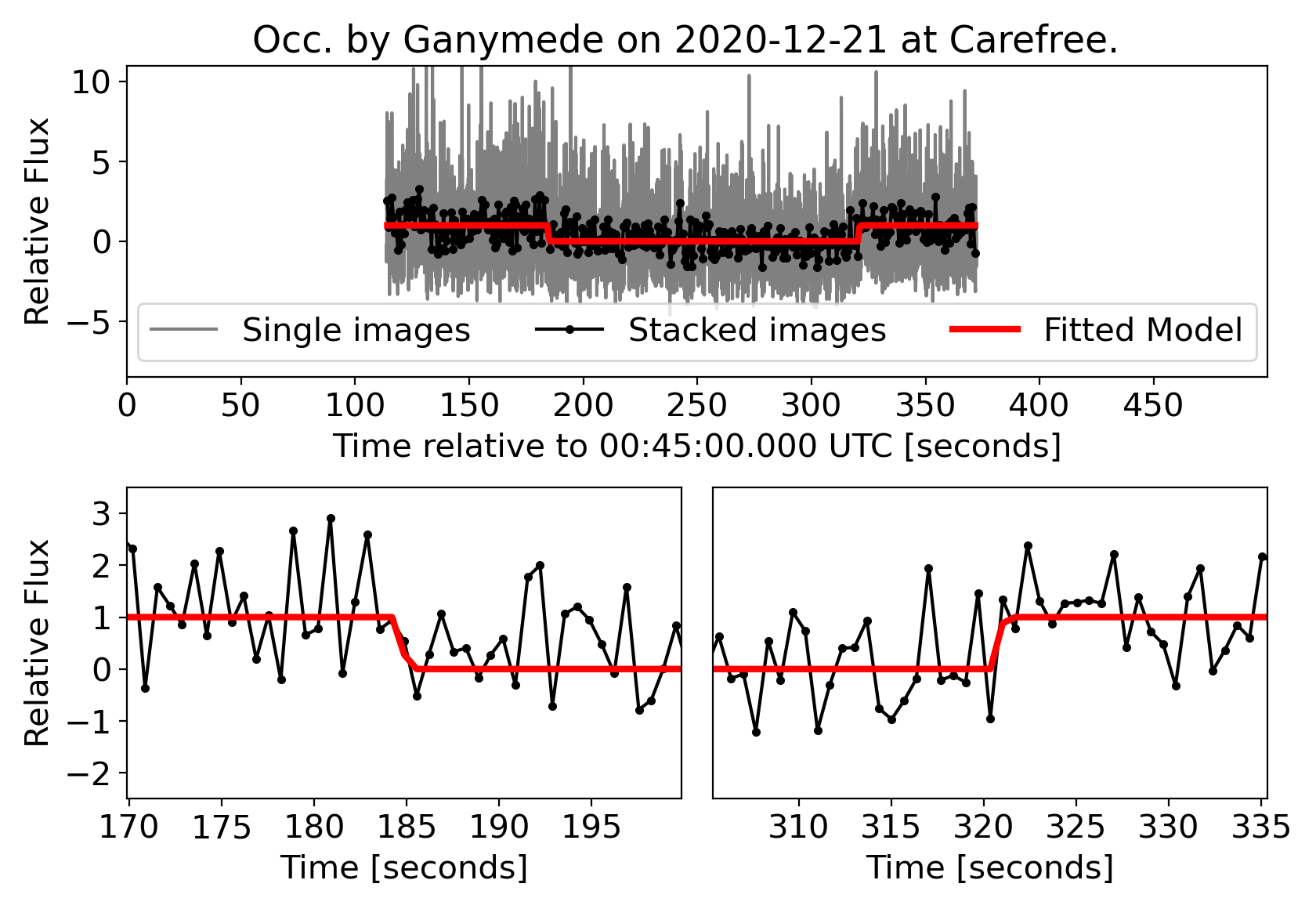}
\caption{Occ. of Ganymede on 2020-12-21 at Carefree by P. Maley.}
\end{figure}               

\begin{figure}[h]
\centering
\includegraphics[width=0.49\textwidth]{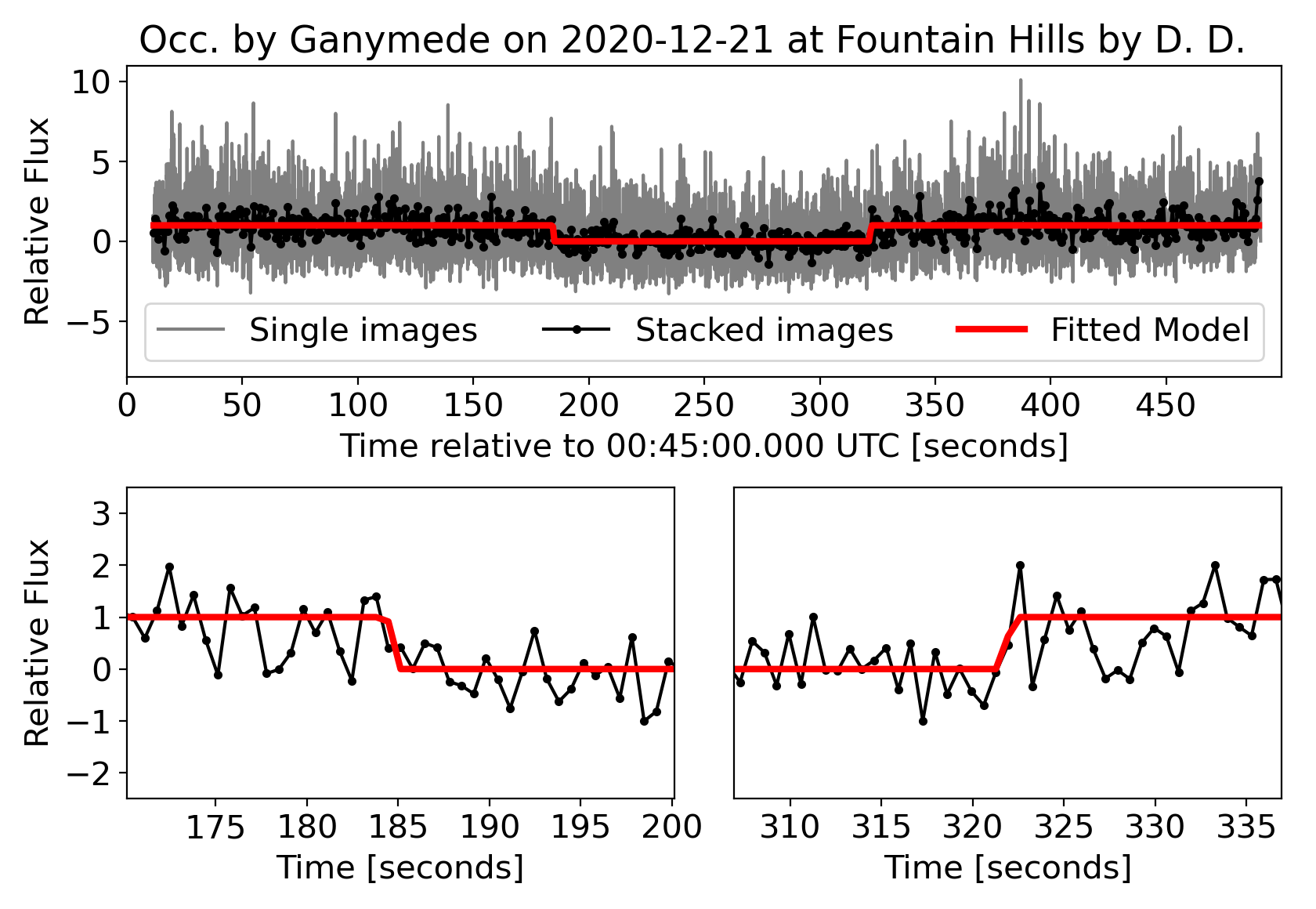}
\caption{Occ. of Ganymede on 2020-12-21 at Fountain Hills by D. Dunham.}
\end{figure}               

\begin{figure}[h]
\centering
\includegraphics[width=0.49\textwidth]{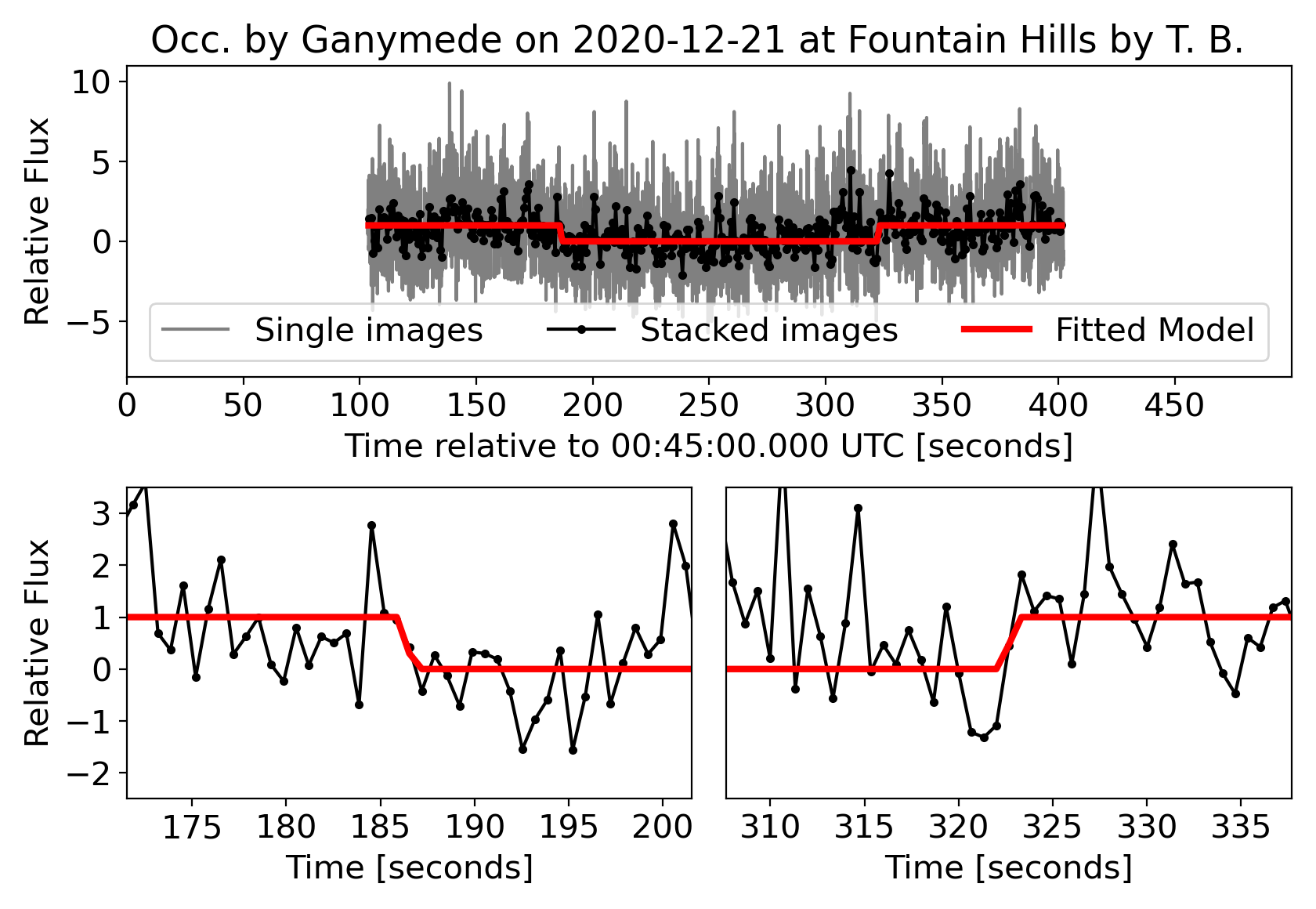}
\caption{Occ. of Ganymede on 2020-12-21 at Fountain Hills by T. Blank.}
\end{figure}               

\begin{figure}[h]
\centering
\includegraphics[width=0.49\textwidth]{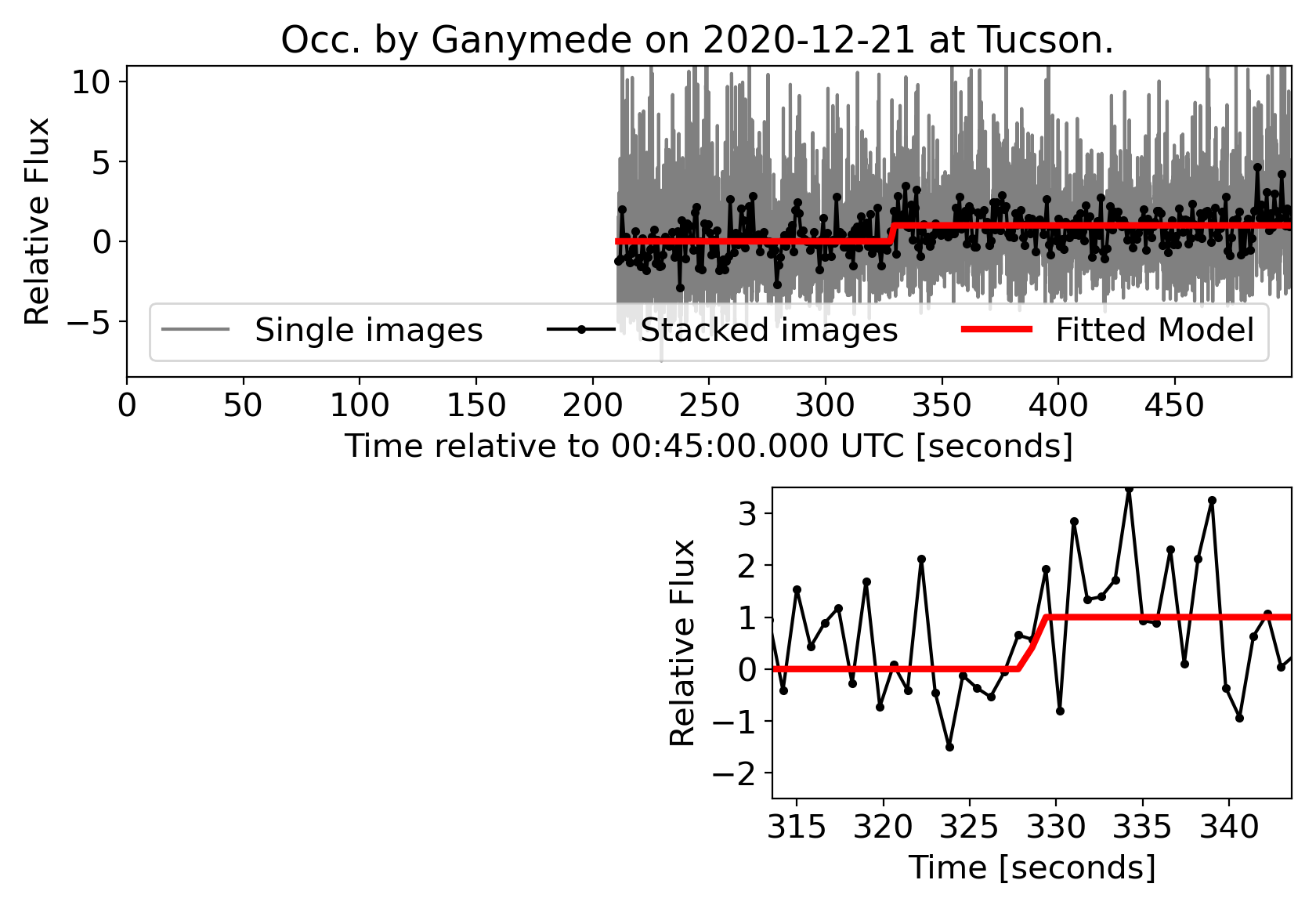}
\caption{Occ. of Ganymede on 2020-12-21 at Tucson by N. Carlson.}
\end{figure}               

\begin{figure}[h]
\centering
\includegraphics[width=0.49\textwidth]{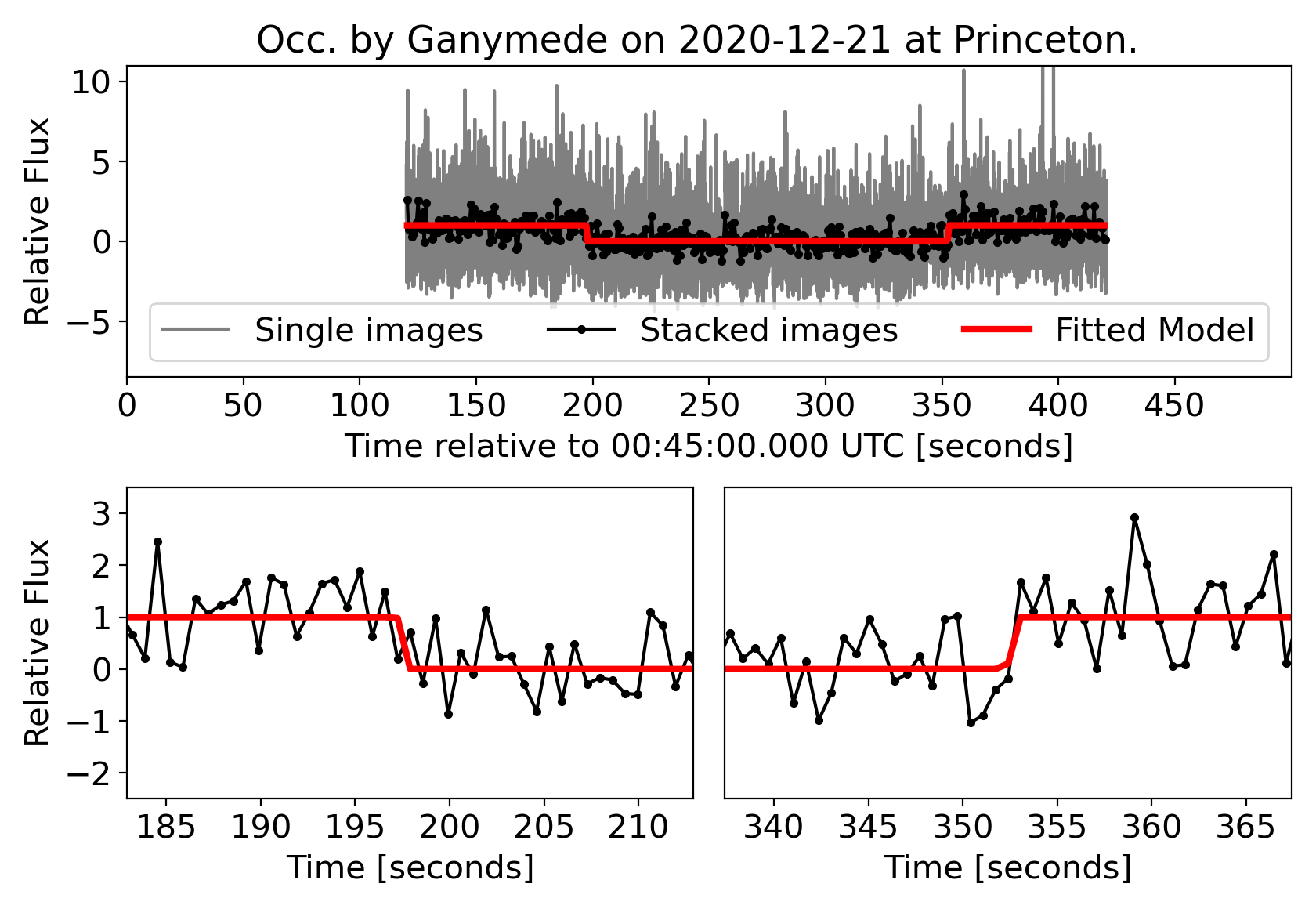}
\caption{Occ. of Ganymede on 2020-12-21 at Princeton by K. Cobble.}
\end{figure}               

\begin{figure}[h]
\centering
\includegraphics[width=0.49\textwidth]{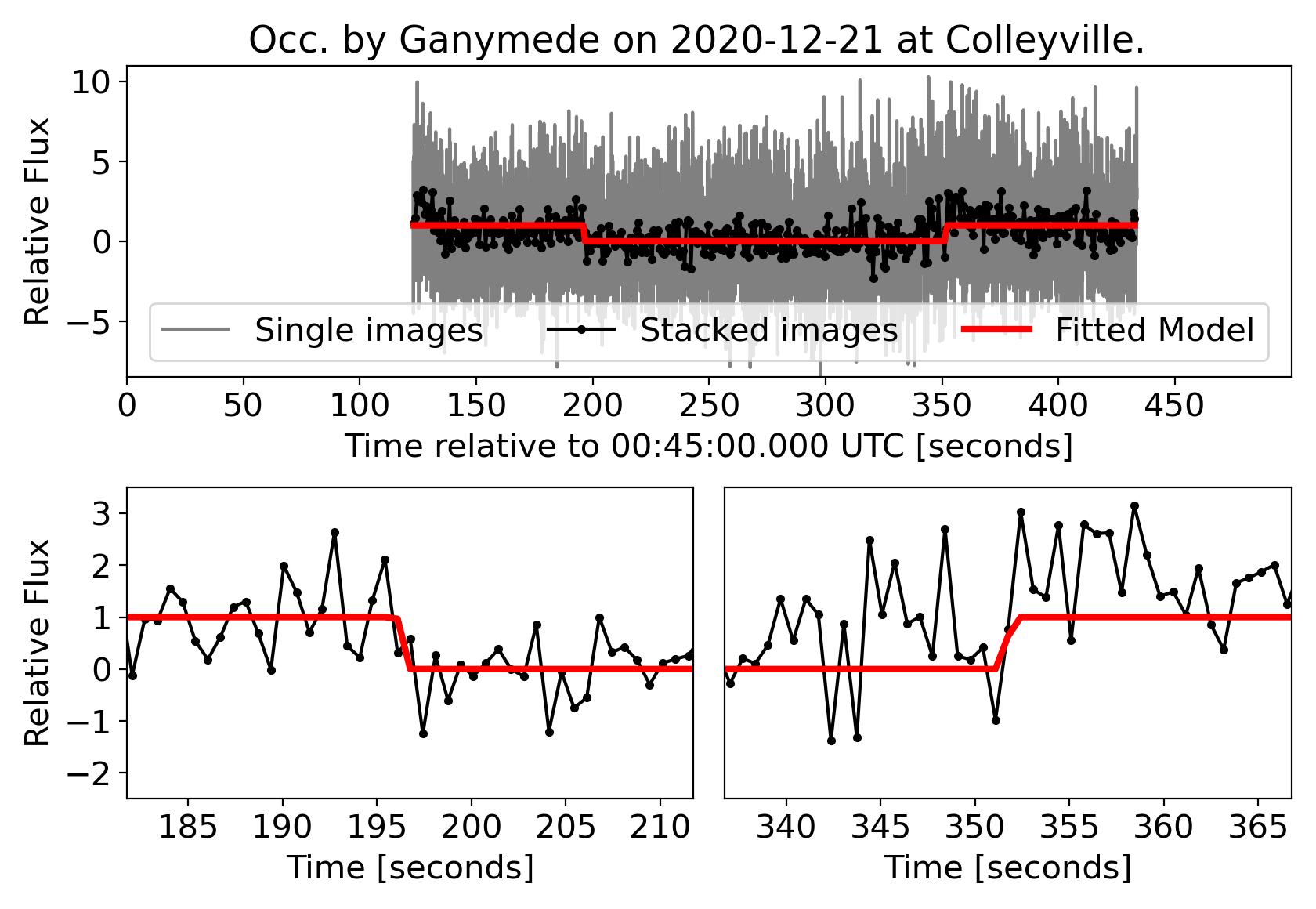}
\caption{Occ. of Ganymede on 2020-12-21 at Colleyville by M. Smith.}
\end{figure}               

\begin{figure}[h]
\centering
\includegraphics[width=0.49\textwidth]{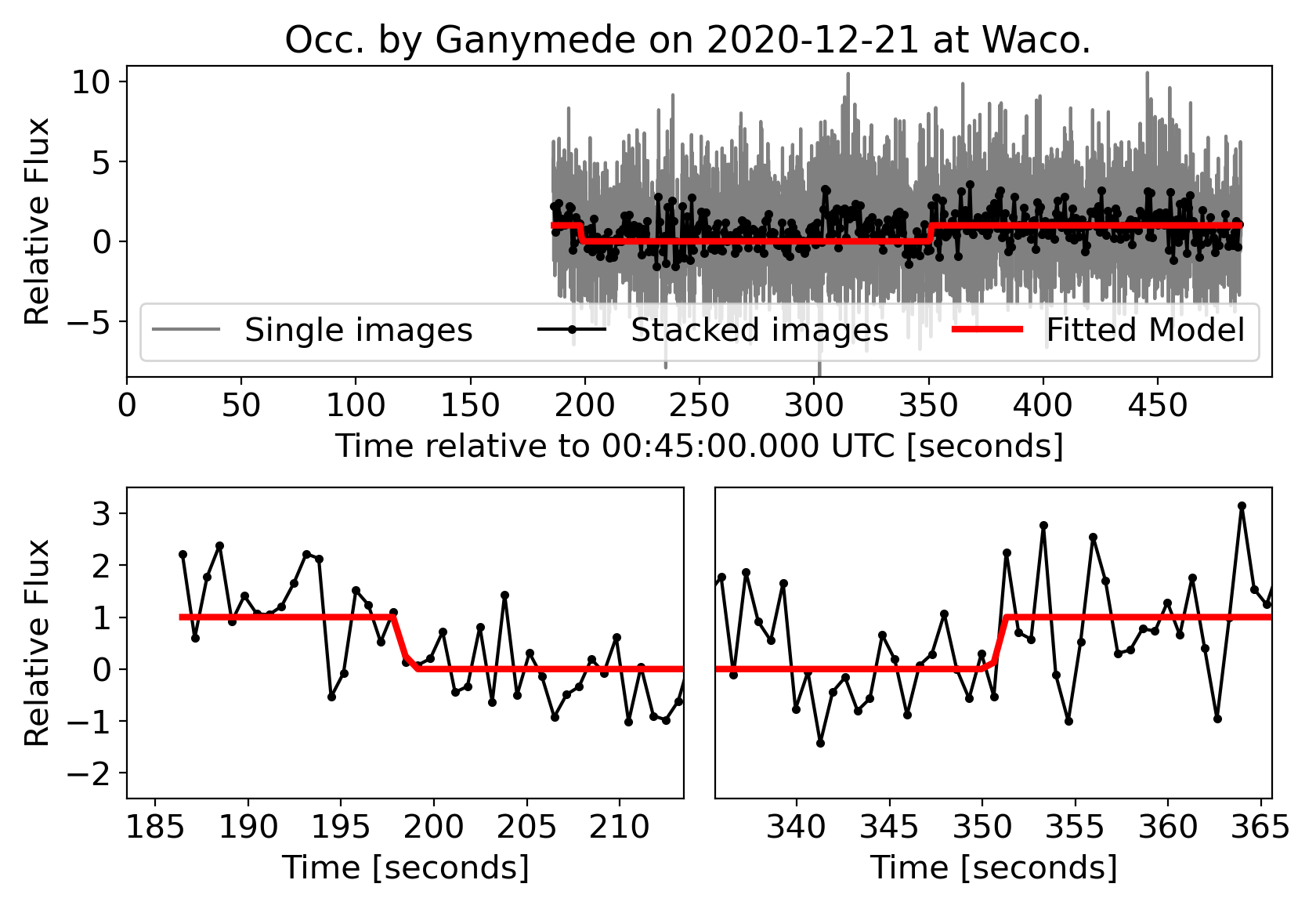}
\caption{Occ. of Ganymede on 2020-12-21 at Waco by D. Eisfeldt.}
\end{figure}               

\begin{figure}[h]
\centering
\includegraphics[width=0.49\textwidth]{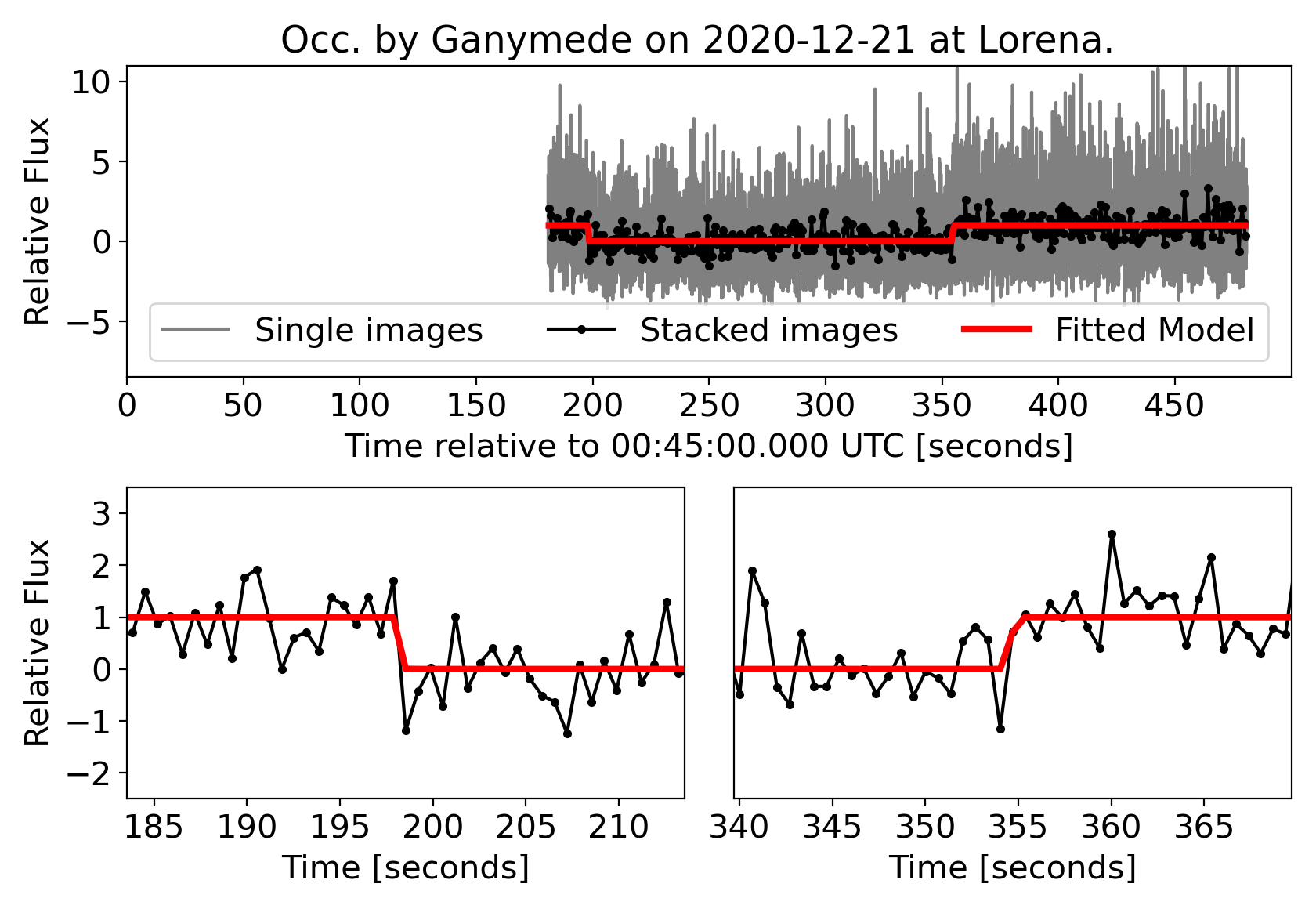}
\caption{Occ. of Ganymede on 2020-12-21 at Lorena by J. Barton.}
\end{figure}               

\begin{figure}[h]
\centering
\includegraphics[width=0.49\textwidth]{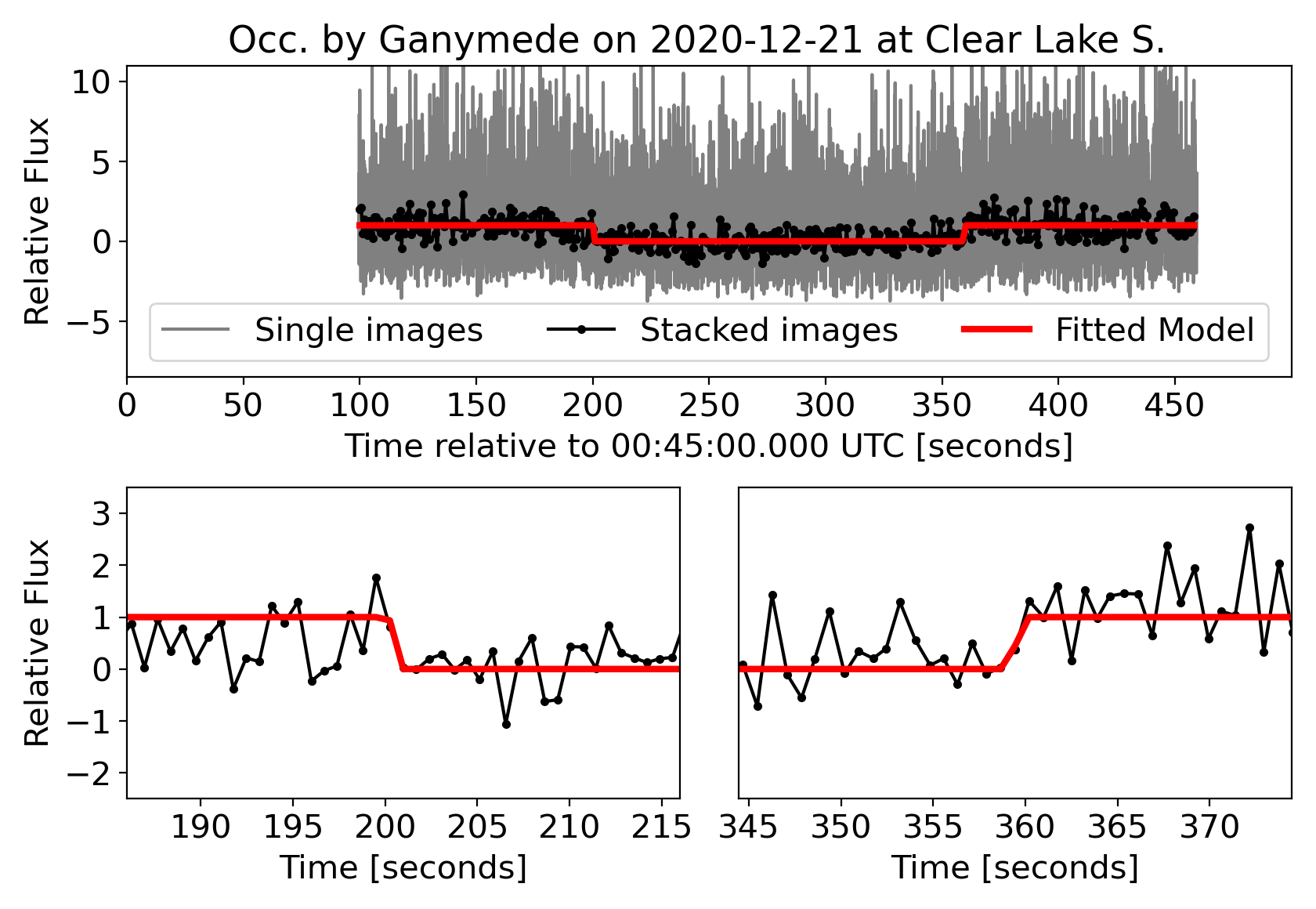}
\caption{Occ. of Ganymede on 2020-12-21 at Clear Lake Shores by P. Stuart.}
\end{figure}               

\begin{figure}[h]
\centering
\includegraphics[width=0.49\textwidth]{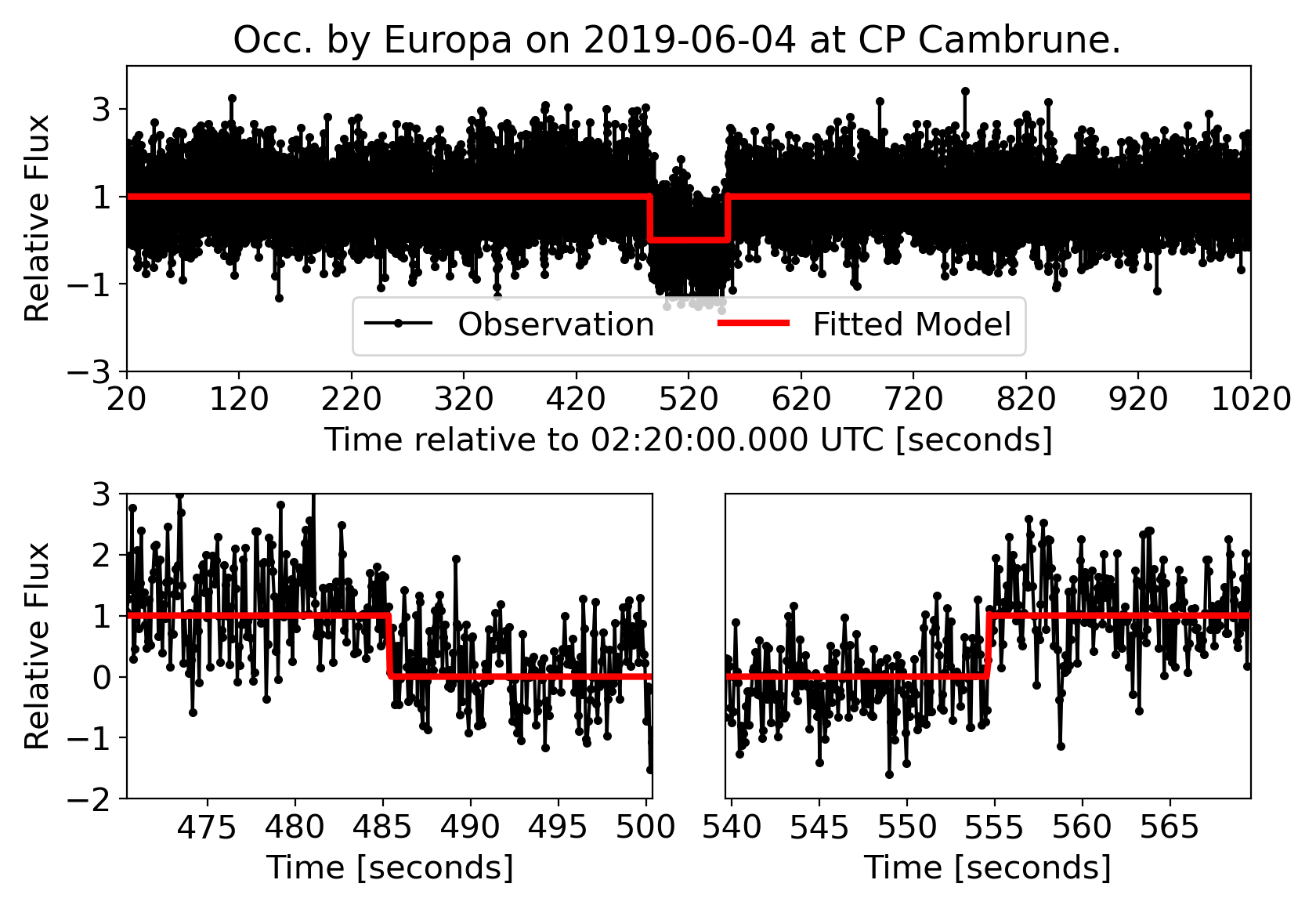}
\caption{Occ. of Europa on 2019-06-04 at CP Cambrune by E. Meza.}
\end{figure}               

\begin{figure}[h]
\centering
\includegraphics[width=0.49\textwidth]{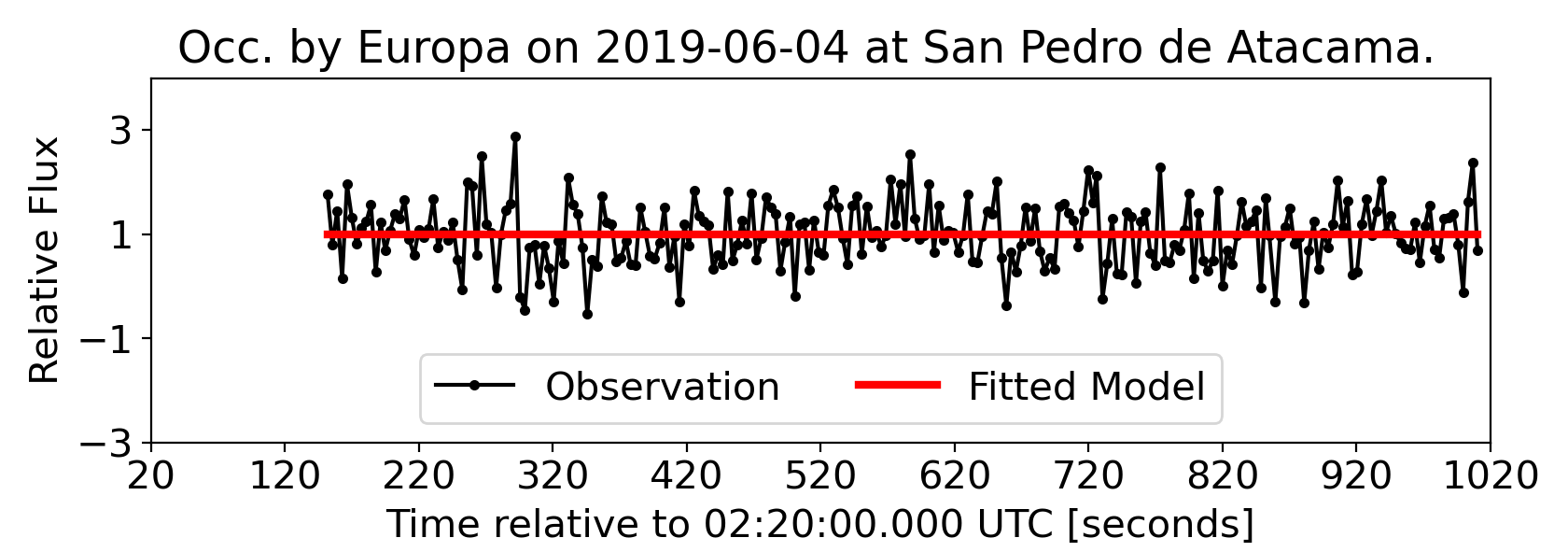}
\caption{Occ. of Europa on 2019-06-04 at San Pedro de Atacama by A. Maury.}
\end{figure}               

\begin{figure}[h]
\centering
\includegraphics[width=0.49\textwidth]{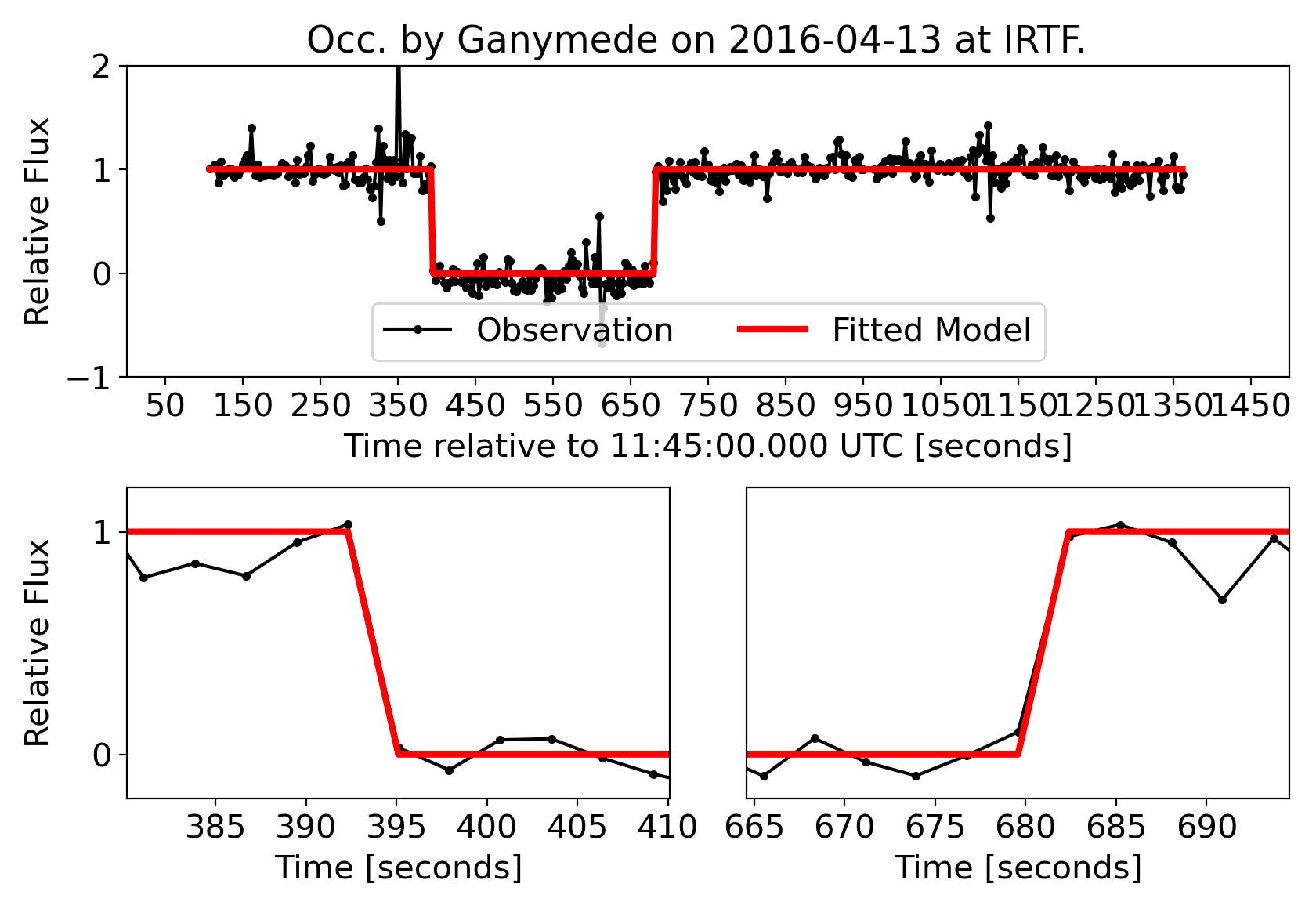}
\caption{Occ. of Ganymede on 2016-04-13 at IRTF by T. Oliva.}
\end{figure}               

\begin{figure}[h]
\centering
\includegraphics[width=0.49\textwidth]{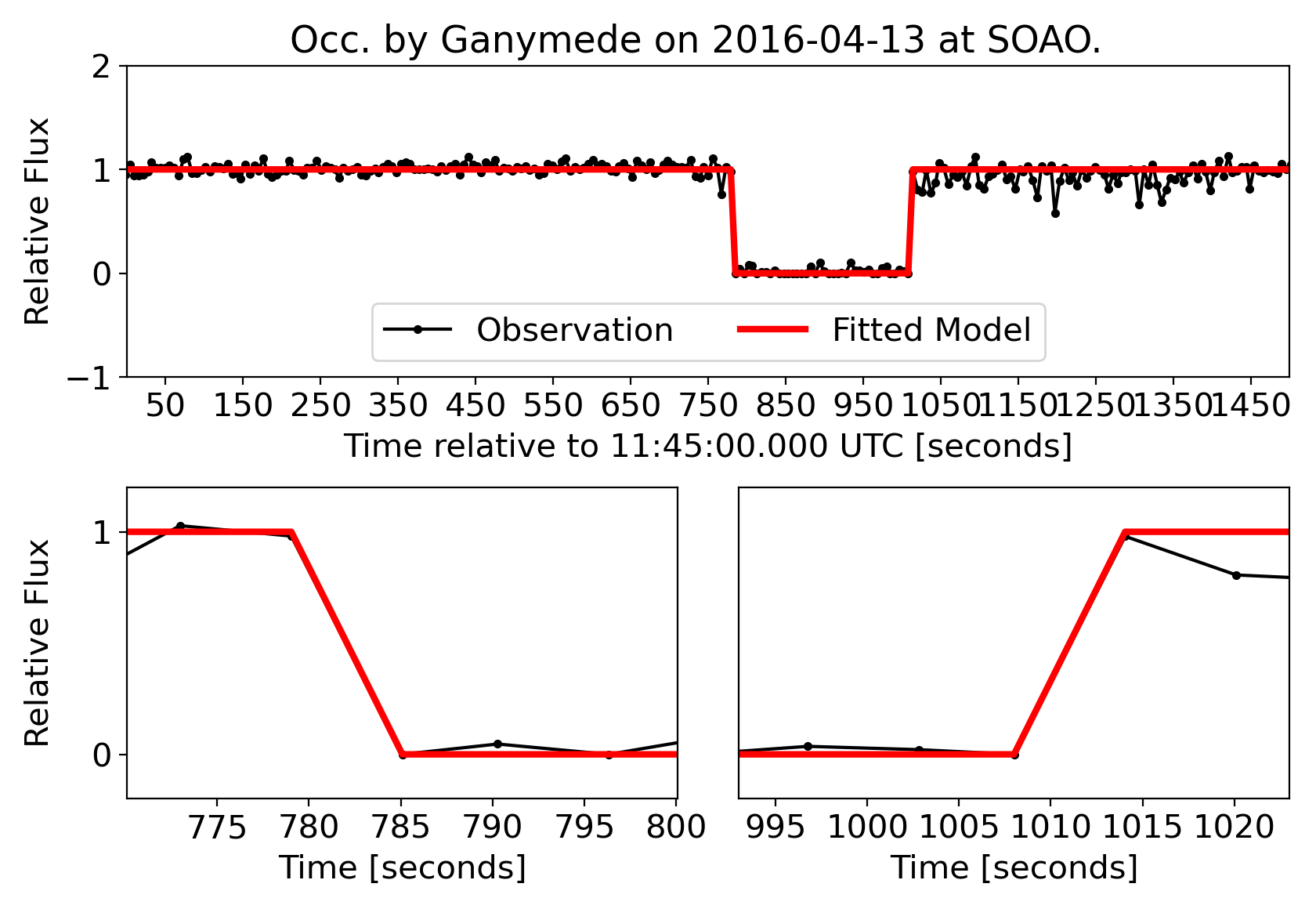}
\caption{Occ. of Ganymede on 2016-04-13 at SOAO by T. C. Hinse and Y. Kim.}
\end{figure}               


\section{Analysis of the Occultation observed with the IRTF SpeX instrument}\label{app:IRTF}

On March 14, 2016, Ganymede occulted a star of magnitude G 6.99. This occultation was favorable to the Northern Pacific Ocean, Japan, South Korea, and Hawaii. At that epoch, \cite{Daversa_2017} organized a campaign to observe this event. 

This observation aimed to search for a signature of Ganymede’s exosphere in the occultation light curve by using the NASA Infrared Telescope Facility (IRTF) on Mauna Kea (Hawaii). At IRTF, both MIT Optical Rapid Imaging System (MORIS) \citep{Gulbis_2011} and SpeX \citep{Rayner_2003} instruments were used, fed by the same optical entrance through a dichroic beam splitter at 0.95 microns. The plan was to MORIS acquire a high-rate sequence of images in the visible range, while SpeX would acquire a series of spectra at a lower rate, covering between 0.9 and 2.5 microns. 

\begin{figure}
    \centering
    \includegraphics[width=\linewidth]{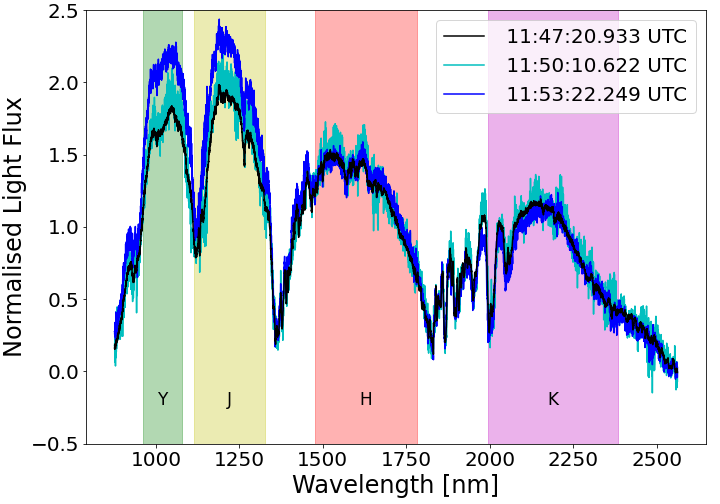}
    \caption{Normalized spectra obtained on the IRTF of the occultation by Ganymede on 2016-04-13. These are the spectra obtained at 11:47:20.933 (in black), 11:50:10.622 (in cyan), and 11:53:22.249 UTC (in blue). We highlight the increase in noise relative to sky fluctuations. This plot also contains the spectral region of the Y, J, H, K bands. For more information, see text.
    }
    \label{fig:spectra}
\end{figure}

The IRTF SpeX data set was obtained between 11:46:47.142 and 12:07:41.851 UTC with a mean cycle of 2.8166 seconds, resulting in 440 spectra where the target star and Ganymede were measured in the same aperture. \autoref{fig:spectra} contains three examples of spectra obtained at three instants. Also, we show in this figure the spectral range of bands: Y (centred at $\lambda_0$ = 1020 nm with a width $\Delta\lambda$ = 120 nm), J ($\lambda_0$ = 1220 nm, $\Delta\lambda$ = 213 nm), H ($\lambda_0$ = 1630 nm, $\Delta\lambda$ = 307 nm) and K ($\lambda_0$ = 2190 nm, $\Delta\lambda$ = 390 nm). We highlight that the instruments' FoV was not large enough to include a reference unocculted body, hence sky fluctuations are the major source of flux variation and noise.

\begin{figure}
    \centering
    \includegraphics[width=\linewidth]{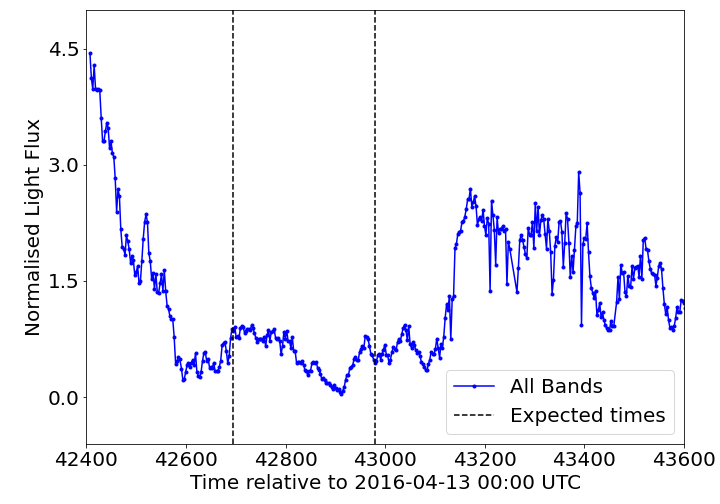}
    \caption{Normalized light curve obtained on the IRTF of the occultation by Ganymede on 2016-04-13. All the spectra were summed for each instant resulting in a light flux. Note that there is a significant flux variation due to sky fluctuations, precluding the detection of the event.
    }
    \label{fig:irtf_lc}
\end{figure}

Apart from the IRTF, this occultation was also observed in the Sobaeksan Optical Astronomy Observatory (SOAO) in South Korea. The data set obtained was analyzed using the standard procedures described in \Autoref{sec:lightcurve}. From the SOAO light curve, we determined dis- and re-appearance times of 11:58:02.028 $\pm$ 2.222 and 12:01:51.064 $\pm$ 2.218. Considering a circle of radius 2631.2 km and the SOAO chord, this would mean that the dis- and re-appearance times of IRTF would be around 11:51:34 and 11:56:19 UTC. However, a direct approach combining the light flux in all the spectral regions to provide a raw light curve shows significant flux variation due to sky fluctuations, precluding the detection of the event, as can be seen in \Autoref{fig:irtf_lc}. However, as seen in \Autoref{fig:spectra}, there is a difference between the spectra outside the expected occultation times (cyan and black lines) and the spectra within the expected occultation times (blue line).

\begin{figure}
    \centering
    \includegraphics[width=\linewidth]{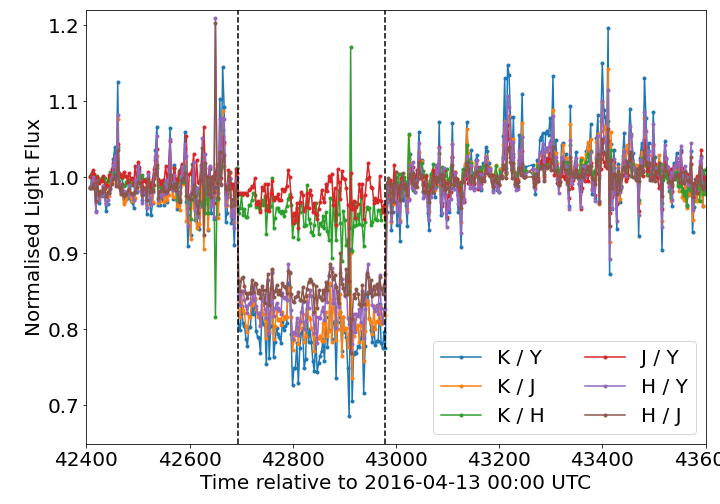}
    \caption{Normalized light curve obtained on the IRTF of the occultation by Ganymede on 2016-04-13. Here we plot the ratio between fluxes for each specific band (Y, J, K, H). The dis- and re-appearance times are clearly captured in four ratios with either Y or J bands at the denominator (H/Y, H/J, K/Y, K/J), while for the other ratios (J/Y, K/H), these times are visually indistinguishable. The dashed vertical lines stand for the expected times. We highlight the agreement between the expected times and the visual drops. 
    }
    \label{fig:band_ratios}
\end{figure}

A different approach is to do the color curves of this data set, as described in the main text. The expected magnitudes of the occulted star are J 5.549, H 5.176, and K 4.908 \citep[NOMAD-1 catalog,][]{NOMAD}. On the other hand, the expected magnitudes of Ganymede in these filters are J 4.045, H 4.145, and K 4.225, based on a V magnitude of 5.045 for the occultation epoch, as obtained using the JPL Horizons service, and the colors provided by \cite{AstroQuanti}. From these values, we can calculate the magnitude drop ($\Delta M_{i}$) for a standard magnitude band $i$ using \autoref{Eq:mag-drop}. 

\begin{equation} \label{Eq:mag-drop}
    \Delta M_i = M_{\circ i} - M_{\star i} + 2.5\log(1+10^{0.4(M_{\star i} - M_{\circ i})}) ~,
\end{equation}
where $M_{\circ i}$ stands for the magnitude of the occulting body and $M_{\star i}$ is the magnitude of the occulted star, both in the band $i$. For example, the occultation in the J band would have a magnitude drop of about 0.24, whereas, in the H band, it would be 0.35. \autoref{fig:band_ratios} shows six possible ratios considering the magnitudes Y, J, H, and K. We call attention to the fact that the occultation can be seen in four of the flux ratios. This difference can be used to determine the occultation instants without a photometric calibration, as we can use one band as a calibrator for another.

Based on the flux ratios, we can fit the dis- or reappearance time using a similar approach as explained in \Autoref{sec:lightcurve}. However, instead of using a standard model, we now use the ratio of two models, one for each band. The effect of the stellar diameter in this modified model would be negligible once it affects both models in the same manner. The fitted times using different flux ratios are equivalent, as can be seen in \autoref{tb:band_ratio}. 

\begin{figure}
    \centering
    \includegraphics[width=1.0\linewidth]{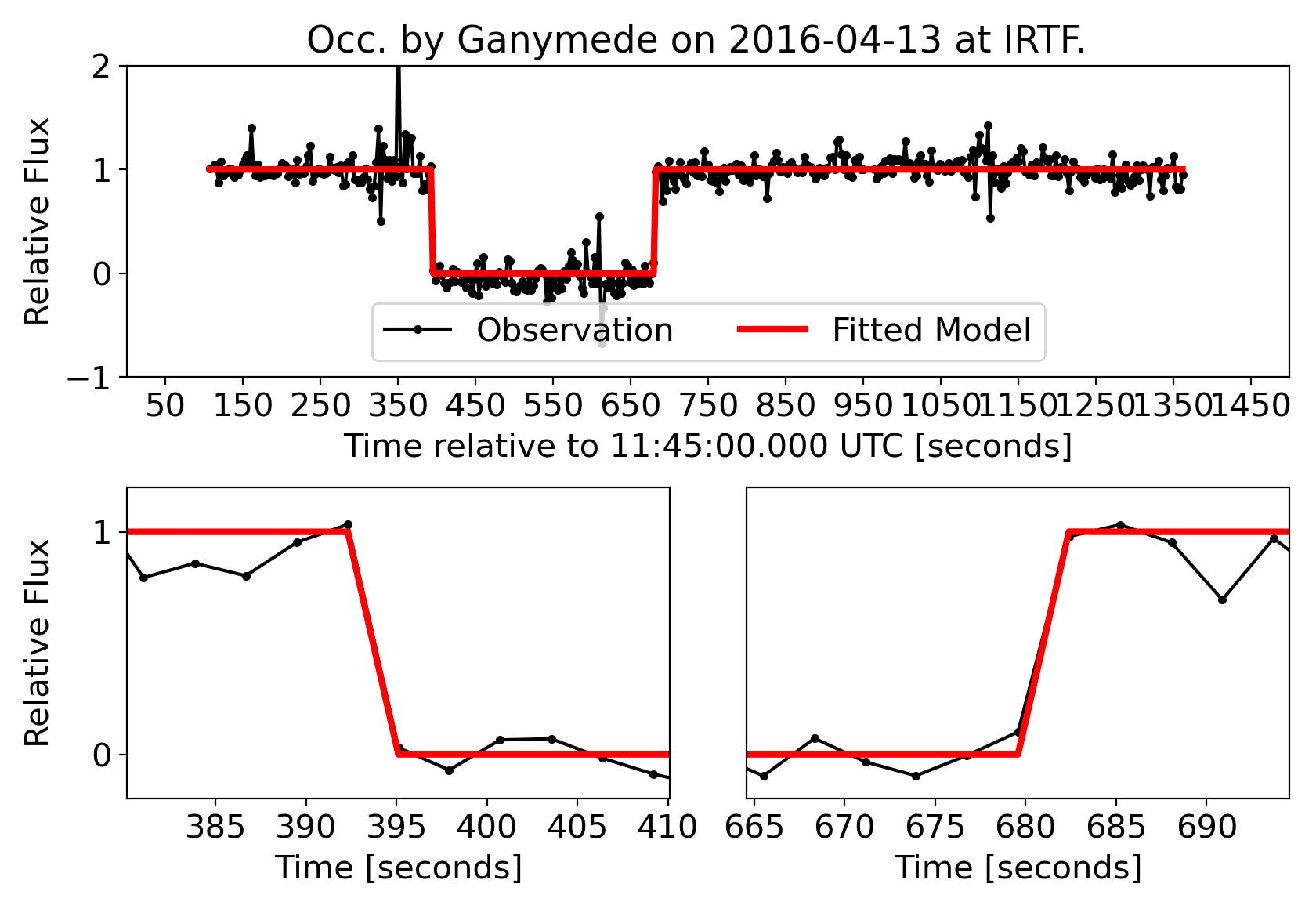}
    \caption{Normalized light curve obtained on the IRTF of the occultation by Ganymede on 2016-04-13. Here we show the color curve relative to the ratio of the bands H/J in black and the fitted modified occultation model in red. The bottom panels contain a zoom-in on 30 seconds centered on the immersion and emersion times.}
    \label{fig:irtf_lc_final}
\end{figure}

\autoref{fig:irtf_lc_final} shows the normalized light curve considering the H/J flux ratio and its best fitted model, here the The total flux from the star and occulting satellite was normalized to unity outside the occultation and to zero within the occultation. We chose to use the times as obtained from the H/J flux ratios together with the SOAO chord in the limb fit, thus obtaining the astrometric result following the procedure described in \Autoref{sec:astrometry} to obtain the result described in \Autoref{sec:obs}.  

\begin{table}
\begin{center}
\caption{Fitted times obtained for each band ratio.}
\vspace{4pt}
\begin{tabular}{ccccc} 
\hline 
\hline
Band ratio & Immersion time UTC & Emersion time UTC & $\chi^2_{pdf}$  \\ 
 & hh:mm:ss.ss (s) & hh:mm:ss.ss (s) &  \\ 
\hline
\hline
\hline
K / Y & 11:51:33.73 (1.23) & 11:56:21.12 (1.23) & 0.915  \\
K / J & 11:51:33.74 (1.23) & 11:56:21.02 (1.26) & 0.921  \\
H / Y & 11:51:33.74 (1.22) & 11:56:21.06 (1.27) & 0.900  \\
H / J & 11:51:33.72 (1.22) & 11:56:20.94 (1.24) & 0.965  \\

\hline
\hline
\end{tabular}\label{tb:band_ratio}
\end{center}
\end{table}

We highlight that colors curves allowed us to determine the occultation instants with high accuracy even without a standard photometric calibrator. That was possible because we are using the relative fluxes, so sky transparency is attenuated, as it affects the analyzed bands in a similar manner, even though the sky transparency was not ideal. This result emphasizes the efficacy of this method, which can be applied to other scientific cases.




\begin{thebibliography}{}

\bibitem[Aksnes \& Franklin(1976)]{Aksnes_1976} Aksnes, K. \& Franklin, F.~A.\ 1976, \aj, 81, 464. doi:10.1086/111908

\bibitem[Aksnes et al.(1984)]{Aksnes_1984} Aksnes, K., Franklin, F., Millis, R., et al.\ 1984, \aj, 89, 280. doi:10.1086/113511

\bibitem[Archinal et al.(2018)]{Archinal_2018} Archinal, B.~A., Acton, C.~H., A'Hearn, M.~F., et al.\ 2018, Celestial Mechanics and Dynamical Astronomy, 130, 22. doi:10.1007/s10569-017-9805-5

\bibitem[Arlot et al.(2014)]{Arlot_2014} Arlot, J.-E., Emelyanov, N., Varfolomeev, M.~I., et al.\ 2014, \aap, 572, A120. doi:10.1051/0004-6361/201423854

\bibitem[Assafin et al.(2011)]{Assafin_2011} Assafin, M., Vieira Martins, R., Camargo, J.~I.~B., et al.\ 2011, Gaia follow-up network for the solar system objects : Gaia FUN-SSO workshop proceedings, 85

\bibitem[Astropy Collaboration et al.(2013)]{astropy} Astropy Collaboration, Robitaille, T.~P., Tollerud, E.~J., et al.\ 2013, \aap, 558, A33

\bibitem[Brozovi{\'c} et al.(2020)]{Brozovic_2020} Brozovi{\'c}, M., Nolan, M.~C., Magri, C., et al.\ 2020, \aj, 159, 149. doi:10.3847/1538-3881/ab7023

\bibitem[Butcher \& Stevens(1981)]{iraf} Butcher, H. \& Stevens, R.\ 1981, Kitt Peak National Observatory Newsletter, No. 16, P. 6, 1981, 16

\bibitem[Butkevich \& Lindegren(2014)]{Butkevich_2014} Butkevich, A.~G. \& Lindegren, L.\ 2014, \aap, 570, A62. doi:10.1051/0004-6361/201424483

\bibitem[Cantat-Gaudin \& Brandt(2021)]{Cantat-Gaudin_2021} Cantat-Gaudin, T. \& Brandt, T.~D.\ 2021, \aap, 649, A124. doi:10.1051/0004-6361/202140807

\bibitem[Carlson et al.(1973)]{Carlson_1973} Carlson, R.~W., Bhattacharyya, J.~C., Smith, B.~A., et al.\ 1973, Science, 182, 53. doi:10.1126/science.182.4107.53


\bibitem[Charnoz et al.(2011)]{Charnoz_2011} Charnoz, S., Crida, A., Castillo-Rogez, J.~C., et al.\ 2011, \icarus, 216, 535. doi:10.1016/j.icarus.2011.09.017

\bibitem[D'Aversa et al.(2017)]{Daversa_2017} D'Aversa, E., Oliva, F., Sindoni, G., et al.\ 2017, European Planetary Science Congress

\bibitem[Cox \& Pilachowski(2000)]{AstroQuanti} Cox, A.~N. \& Pilachowski, C.~A.\ 2000, Physics Today, 53, 77. doi:10.1063/1.1325201

\bibitem[Crida \& Charnoz(2012)]{Crida_2012} Crida, A. \& Charnoz, S.\ 2012, Science, 338, 1196. doi:10.1126/science.1226477

\bibitem[Dirkx et al.(2016)]{Dirkx_2016} Dirkx, D., Lainey, V., Gurvits, L.~I., et al.\ 2016, \planss, 134, 82. doi:10.1016/j.pss.2016.10.011

\bibitem[Dirkx et al.(2017)]{Dirkx_2017} Dirkx, D., Gurvits, L.~I., Lainey, V., et al.\ 2017, \planss, 147, 14. doi:10.1016/j.pss.2017.09.004

\bibitem[Emelyanov(2009)]{Emelyanov_2009} Emelyanov, N.~V.\ 2009, \mnras, 394, 1037. doi:10.1111/j.1365-2966.2009.14398.x

\bibitem[Gaia Collaboration et al.(2016a)]{GAIA2016a} Gaia Collaboration, Prusti, T., de Bruijne, J.~H.~J., et al.\ 2016, \aap, 595, A1. doi:10.1051/0004-6361/201629272

\bibitem[Gaia Collaboration et al.(2016b)]{GAIA2016b} Gaia Collaboration, Brown, A.~G.~A., Vallenari, A., et al.\ 2016, \aap, 595, A2. doi:10.1051/0004-6361/201629512

\bibitem[Gaia Collaboration et al.(2018)]{GAIA2018} Gaia Collaboration, Brown, A.~G.~A., Vallenari, A., et al.\ 2018, \aap, 616, A1. doi:10.1051/0004-6361/201833051

\bibitem[Gaia Collaboration et al.(2021)]{GAIA2021} Gaia Collaboration, Brown, A.~G.~A., Vallenari, A., et al.\ 2021, \aap, 649, A1. doi:10.1051/0004-6361/202039657

\bibitem[Gomes-J{\'u}nior et al.(2016)]{Gomes-Junior_2016} Gomes-J{\'u}nior, A.~R., Assafin, M., Beauvalet, L., et al.\ 2016, \mnras, 462, 1351. doi:10.1093/mnras/stw1738

\bibitem[Gomes-J{\'u}nior et al.(2022)]{sora} Gomes-J{\'u}nior, A.~R., Morgado, B.~E., Benedetti-Rossi, G., et al.\ 2022, arXiv:2201.01799

\bibitem[Gulbis et al.(2011)]{Gulbis_2011} Gulbis, A.~A.~S., Bus, S.~J., Elliot, J.~L., et al.\ 2011, \pasp, 123, 461. doi:10.1086/659636

\bibitem[Herald et al.(2020)]{Herald_2020} Herald, D., Gault, D., Anderson, R., et al.\ 2020, \mnras, 499, 4570. doi:10.1093/mnras/staa3077

\bibitem[Hubbard \& van Flandern(1972)]{Hubbard_1972} Hubbard, W.~B. \& van Flandern, T.~C.\ 1972, \aj, 77, 65. doi:10.1086/111246

\bibitem[Kiseleva et al.(2008)]{Kiseleva_2008} Kiseleva, T.~P., Izmailov, I.~S., Kiselev, A.~A., et al.\ 2008, \planss, 56, 1908. doi:10.1016/j.pss.2008.02.024

\bibitem[Lainey et al.(2009)]{Lainey_2009} Lainey, V., Arlot, J.-E., Karatekin, {\"O}., et al.\ 2009, \nat, 459, 957. doi:10.1038/nature08108

\bibitem[Lainey et al.(2012)]{Lainey_2012} Lainey, V., Karatekin, {\"O}., Desmars, J., et al.\ 2012, \apj, 752, 14. doi:10.1088/0004-637X/752/1/14

\bibitem[Lainey et al.(2017)]{Lainey_2017} Lainey, V., Jacobson, R.~A., Tajeddine, R., et al.\ 2017, \icarus, 281, 286. doi:10.1016/j.icarus.2016.07.014

\bibitem[Morgado et al.(2016)]{Morgado_2016} Morgado, B., Assafin, M., Vieira-Martins, R., et al.\ 2016, \mnras, 460, 4086. doi:10.1093/mnras/stw1244

\bibitem[Morgado et al.(2019a)]{Morgado_2019a} Morgado, B., Vieira-Martins, R., Assafin, M., et al.\ 2019, \mnras, 482, 5190. doi:10.1093/mnras/sty3040

\bibitem[Morgado et al.(2019b)]{Morgado_2019b} Morgado, B., Benedetti-Rossi, G., Gomes-J{\'u}nior, A.~R., et al.\ 2019, \aap, 626, L4. doi:10.1051/0004-6361/201935500

\bibitem[Morgado et al.(2019c)]{Morgado_2019c} Morgado, B., Vieira-Martins, R., Assafin, M., et al.\ 2019, \planss, 179, 104736. doi:10.1016/j.pss.2019.104736

\bibitem[Nimmo et al.(2007)]{Nimmo_2007} Nimmo, F., Thomas, P.~C., Pappalardo, R.~T., et al.\ 2007, \icarus, 191, 183. doi:10.1016/j.icarus.2007.04.021

\bibitem[Press et al.(1992)]{Numerical_Recipes} Press, W.~H., Teukolsky, S.~A., Vetterling, W.~T., et al.\ 1992, Cambridge: University Press, |c1992, 2nd ed.

\bibitem[Rayner et al.(2003)]{Rayner_2003} Rayner, J.~T., Toomey, D.~W., Onaka, P.~M., et al.\ 2003, \pasp, 115, 362. doi:10.1086/367745

\bibitem[Saquet et al.(2018)]{Saquet_2018} Saquet, E., Emelyanov, N., Robert, V., et al.\ 2018, \mnras, 474, 4730. doi:10.1093/mnras/stx2957

\bibitem[Sicardy et al.(2011)]{Sicardy_2011} Sicardy, B., Ortiz, J.~L., Assafin, M., et al.\ 2011, \nat, 478, 493. doi:10.1038/nature10550

\bibitem[Thomas et al.(1997)]{Thomas_1997} Thomas, P.~C., Simonelli, D., Burns, J., et al.\ 1997, Lunar and Planetary Science Conference

\bibitem[White et al.(2014)]{White_2014} White, O.~L., Schenk, P.~M., Nimmo, F., et al.\ 2014, Journal of Geophysical Research (Planets), 119, 1276. doi:10.1002/2013JE004591

\bibitem[Zacharias et al.(2004)]{NOMAD} Zacharias, N., Monet, D.~G., Levine, S.~E., et al.\ 2004, \aas

\bibitem[Zubarev et al.(2015)]{Zubarev_2015} Zubarev, A., Nadezhdina, I., Oberst, J., et al.\ 2015, \planss, 117, 246. doi:10.1016/j.pss.2015.06.022

\end{thebibliography}
\end{document}